\documentclass[12pt]{article}%
\usepackage{amsmath}
\usepackage{amsfonts}
\usepackage{amssymb}
\usepackage{graphicx}
\usepackage{float}
\usepackage{xcolor}
\usepackage{color}
\usepackage{geometry}
\usepackage{natbib}%
\usepackage{amsthm}
\usepackage{algorithm}
\usepackage{algorithmic}
\usepackage{hyperref}

\numberwithin{equation}{section}

\numberwithin{equation}{section}

\newcommand{\be}{\begin{equation}}
\newcommand{\ee}{\end{equation}}

\newcommand{\bq}{\begin{eqnarray}}
\newcommand{\eq}{\end{eqnarray}}

\setcitestyle{authoryear,round}

\begin{document}

\title{Navigating Uncertainty in ESG Investing \footnote{*We thank Marcel Oestreich, Amr ElAlfy, Chengguo Weng, and Johnny Li, whose constructive comments and suggestions have
improved both the substance and presentation of the paper. The usual disclaimer applies. We also thank the organizers and the participants at Fields - Institute's Mathematics for Climate Change (MfCC) Network \& Waterloo Institute for Complexity and Innovation (WICI): Math for Complex Climate Challenges Workshop at the University of Waterloo, Waterloo, May 1 - 4, 2023, the 26th International Congress on Insurance: Mathematics and Economics at the Department of Actuarial Mathematics and Statistics at Heriot-Watt University, Edinburgh from July 4 - 7, 2023, and the 58th Actuarial Research Conference (ARC) at Drake University in Des Moines, Iowa, July 30 - August 2, 2023.}}
\author{Jiayue Zhang\thanks{Department of Statistics \& Actuarial Science, University of Waterloo (j857zhan@uwaterloo.ca)}
\and Ken Seng Tan\thanks{Division of Banking and Finance, Nanyang Technological University (kenseng.tan@ntu.edu.sg)} 
\and Tony S. Wirjanto \thanks{Department of Statistics \& Actuarial Science, University of Waterloo (twirjanto@uwaterloo.ca)} \thanks{School of Accounting \& Finance, University of Waterloo (twirjanto@uwaterloo.ca)}
\and 
Lysa Porth \thanks{Gordon S. Lang School of Business and Economics, University of Guelph (lporth@uoguelph.ca)}
}
\date{{\small September 2025 }}
\maketitle
\begin{abstract}
\noindent The widespread confusion among investors regarding Environmental, Social, and Governance (ESG) rankings assigned by rating agencies has underscored a critical issue in sustainable investing. To address this uncertainty, our research has devised methods that not only recognize this ambiguity but also offer tailored investment strategies for different investor profiles. By developing ESG ensemble strategies and integrating ESG scores into a Reinforcement Learning (RL) model, we aim to optimize portfolios that cater to both financial returns and ESG-focused outcomes. Additionally, by proposing the Double-Mean-Variance model, we classify three types of investors based on their risk preferences. We also introduce ESG-adjusted Capital Asset Pricing Models (CAPMs) to assess the performance of these optimized portfolios. Ultimately, our comprehensive approach provides investors with tools to navigate the inherent ambiguities of ESG ratings, facilitating more informed investment decisions.




\end{abstract}

\makeatletter

\makeatother
\baselineskip 18pt

\newpage

\section{Introduction}
The integration of Environmental, Social, and Governance (ESG) factors into capital markets represents a significant evolution in financial analysis and investment strategies. As global markets increasingly recognize the importance of sustainable practices, ESG factors have become critical indicators in evaluating the long-term viability and ethical positioning of enterprises. However, the assimilation of ESG criteria into capital markets is not without its complexities. A primary concern is the inconsistency and perceived unreliability of ESG data. Empirical evidence, such as that provided by
\cite{berg2022aggregate}, 
indicates notable discrepancies among major ESG rating agencies, with sample correlation coefficients as low as 0.54 on average. This contrasts sharply with the high consistency found in credit ratings by established firms like Moody's and Standard \& Poor's where sample autocorrelation coefficients are 0.92 on average. Such heterogeneity presents challenges for investors aiming at balancing both financial and ESG-centric return objectives. This prevailing heterogeneity around ESG ratings inevitably leads to a widespread confusion among investors who are striving to achieve both financial and ESG factor return objectives. 

There are varying perspectives on how this confusion impacts investors. Some believe that the ambiguity could allow investors to play a more active role in their asset decisions. Conversely, others posit that the absence of ambiguity could result in a complete integration of ESG factors into the market, potentially diminishing the profitability of trading strategies. Efforts to tackle this challenge are underway, exemplified by initiatives such as the ongoing research conducted by the Aggregate Confusion Project within MIT's Sloan Sustainability Initiative.

The objective of this paper is to propose methods for asset allocation for investors amidst the prevalent ESG rating confusion with three major applications in
finance.

Firstly, we incorporate the ESG scores explicitly in the reward function of a conventional reinforcement learning (RL) model, which is one of the tookits in machine learning, to analyze the effect of widespread confusion caused by the heterogeneity in the ESG ratings on the coherence of investment strategies. In response to this confusion, the approach undertaken in this paper does not rely on just a single ESG rating system but introduces several ensemble strategies tailored to varying investor profiles regarding risk and ambiguity preferences.

Secondly, we present a Double-Mean-Variance (DMV) model to categorize investors based on their stance on ESG and their comfort level with rating uncertainties. The three types of investors are (i) those who are indifferent to ESG, (ii) those who prefer ESG but are not affected by uncertainty with respect to the ESG scores, and (iii) those who prefer ESG and are affected by uncertainty with respect to the ESG scores. In the calibration exercise, we propose a refined RL model by setting its reward function to be the specified DMV model under different sources of uncertainty associated with ESG ratings. 

Thirdly, we introduce an ESG-modified Capital Asset Pricing Model (CAPM) to assess the performance of the optimized portfolio for different types of ESG-oriented investors.

The remaining parts of this paper are organized as follows. Section \ref{2.2} confirms the observed heterogeneity in ESG ratings, using samples from four prominent rating agencies. The implications of these discrepancies on portfolio selection are discussed, adopting an RL framework for calibration. In Section \ref{2.3}, instead of a singular focus on one agency's ESG ratings, we introduce strategies to amalgamate ratings from the four key agencies. This approach aims at accommodating investors' varying risk tolerance and perspectives on ESG rating ambiguity, more precisely, uncertainty. In Section \ref{2.4}, we showcase an ESG-focused portfolio model that integrates MV preferences at the same time. We detail a calibration exercise, where ESG scores are directly infused into the RL model's reward function, utilizing the outlined DMV model. Section \ref{2.5} builds upon the DMV model's findings and designs ESG-adapted CAPMs. These cater to investors prioritizing ESG, considering both the presence and absence of uncertainties in ESG ratings. This section proposes a refined CAPM framework to evaluate the efficacy of the resulting optimized portfolios. Section \ref{2.6} provides some concluding remarks for this paper.
\section{Heterogeneity in ESG ratings}
\label{2.2}

A key challenge in ESG investing is the substantial disagreement across ESG rating providers, even when evaluating the same firm. \cite{berg2022aggregate} quantify this divergence and trace it to differences in indicator selection, aggregation weights, and transparency levels. Building on this, \cite{christensen2022corporate} find that greater ESG disclosure actually exacerbates rating disagreement, rather than resolving it, because raters interpret ESG outcome metrics differently. Notably, they show that rating divergence is more pronounced for “outcome” measures (e.g., emissions level) than for “policy” disclosures (e.g., climate pledges), which has implications for how ratings propagate into volatility and capital market behavior. In parallel, ESG ratings may also fail to reflect investor-specific preferences. \cite{assaf2024esg} survey retail investors and find that individuals differ significantly in how they value the E, S, and G components—some favor environmental metrics, others emphasize social or governance factors. They argue that a single ESG score cannot fully represent the multidimensional preferences of investors, especially when regulators require alignment between investment products and client sustainability goals. These findings support the need for customizable, preference-aware ESG integration frameworks, such as the ensemble methods we propose in this paper.

To underscore the contribution of our analysis in this paper, we begin by revisiting the results presented by \cite{berg2022aggregate}. The ESG data used in our study are acquired from four major rating agencies: SA (Sustainalytics), RobecoSAM (S\&P Global), Asset4 (Refinitiv), and MSCI. The data from these rating agencies are downloaded from the Thomson Reuters Eikon database (for Asset4) and Bloomberg (for RobecoSAM, Sustainalytics, and MSCI scores) database respectively. Although the majority of our data originates from 2020, any unavailable updates for that year default to the 2019 data. Additionally, our analysis integrates ESG scores for the Dow Jones 30 stocks derived from the aforementioned raters.

ESG rating agencies can differ in terms of their sample coverage and rating scale. In particular, Asset4, Sustainalytics, and RobecoSAM apply a scale from 0 to 100, MSCI uses a seven-tier rating scale from the best (AAA) to the worst (CCC). 

To harmonize different ESG rating scales across agencies, MSCI's scale is transformed into a $0-100$ format by dividing the scale into seven equal intervals and assigning the average value of each segment to represent the respective grades. This standardization enables consistent treatment of ESG scores in reinforcement learning portfolio optimization, where numerical operations are essential. While such transformation assumes equal distances between adjacent rating levels—an approximation that may not fully reflect the underlying nuances—it is a practical approach supported by industry standards. For instance, MSCI itself converts its internal ESG fund ratings into a numerical “ESG Quality Score” ranging from 0 to 10 and then maps these scores into letter grades, assigning fixed intervals to each (e.g., AAA corresponds to 8.571–10.0, CCC to 0–1.429), please see \cite{msci_esg_ratings}. Our linear transformation into a 0–100 scale is conceptually aligned with this logic and ensures comparability with other providers who already use a 0–100 scale (e.g., Sustainalytics, RobecoSAM, Asset4). Additionally, since our analyses focus on relative performance comparisons (e.g., Sharpe ratio rankings across ESG raters and ensemble methods), the precise shape of the original MSCI mapping is less likely to materially affect the conclusions. Nonetheless, this transformation is acknowledged as a simplification and a potential source of granularity loss.

Table~\ref{tab1} shows the sample correlation of ESG ratings based on Pearson Correlation Coefficients, echoing the findings of \cite{berg2022aggregate}. The highest sample correlation coefficient of the ESG ratings is only 0.504. There are even several pairs of ratings that have negative coefficients. This leads to the prevailing issue of ESG confusion. See \cite{vxw2024} for the use of various correlation and distance measures on corporate ESG ratings and the use of clustering techniques to identify groups of firms in the sample with similar ESG profiles.

\begin{table}[h!]
\begin{tabular}{|l|l|l|l|l|}
\hline & \textbf{RebecoSAM} & \textbf{SA} & \textbf{MSCI} & \textbf{Asset4} \\
\hline
\hline \textbf{RebecoSAM} & 1.0000 & -0.1591 & 0.4153 & 0.5041 \\
\textbf{SA} & -0.1591 & 1.0000 & -0.3387 & 0.1826 \\
\textbf{MSCI} & 0.4153 & -0.3387 & 1.0000 & 0.3139 \\
\textbf{Asset4} & 0.5041 & 0.1826 & 0.3139 & 1.0000 \\
\hline
\end{tabular}
\centering
\caption{The Sample Correlation Matrix of the Four ESG Raters}
\label{tab1}
\end{table}
\subsection{Heterogeneity of ESG raters in portfolio selection}

To better understand the impact of reported heterogeneity in ESG ratings on investors’ asset allocation, we calibrate a portfolio selection process using an explicit RL framework.. 

\subsubsection{Brief introduction to RL}

RL serves as a pivotal paradigm for training agents to make sequential decisions in dynamic environments, as explained in \cite{sutton2018reinforcement} and recently reviewed in \cite{hambly2023recent}. Specifically, RL agents interact with their environment, receiving observations of the current state and rewards or penalties for their actions. The agents then learn to choose actions that maximize the expected cumulative reward, using trial-and-error exploration (of uncharted territory) and exploitation (of current knowledge), as highlighted in \cite{kaelbling1996reinforcement}. This optimization process is crucial in scenarios where a sequential decision-making process takes place. As a result, RL algorithms can be used to explore different investment strategies and learn which ones are effective and which ones are not. RL learns optimal portfolio allocation strategies by continuously interacting with market data and making decisions over time. This approach accounts for the evolving nature of financial markets, where historical patterns may not always hold. RL's capacity to balance exploration and exploitation allows it to uncover potentially profitable investment opportunities while minimizing risk. Several prior studies have found that RL algorithms can outperform traditional investment strategies, particularly in volatile or complex markets. For example, \cite{jiang2017deep} found that an RL algorithm was able to outperform traditional portfolio selection strategies with a high commission rate of 0.25\% in the backtests. \cite{li2021finrl} present a FinRL-Podracer framework to accelerate the development pipeline of deep reinforcement learning (DRL)-driven trading strategy and to improve both trading performance and training efficiency.
A foundational concept in RL is known as Markov Decision Process (MDP), which provides a framework for modeling decision-making in a stochastic environment. The value of each state within an MDP is ascertained through a Bellman equation, which is a fundamental equation that governs a trade-off between immediate rewards and future rewards. The Bellman equation is expressed as:
\begin{equation}
\label{bellman}
V(s)=\max _{a \in A(s)} \sum_{s^{\prime}} P\left(s^{\prime} \mid s, a\right)\left(R\left(s, a, s^{\prime}\right)+\gamma V\left(s^{\prime}\right)\right)
\end{equation}
Here, the symbol $V(s)$ represents the value of state $s$, while $P(s' | s, a)$ denotes the probability of transitioning from state $s$ to state $s'$ when action $a$ is taken. The term $R(s, a, s')$ corresponds to the reward obtained by taking action $a$ in state $s$ and transitioning to state $s'$. The parameter $\gamma$, residing in the interval $[0, 1]$, represents the discount factor that modulates the impact of future rewards, balancing the level of exploration and exploitation in decision-making processes. The objective of the agent is to maximize the cumulative reward, which is accomplished by selecting the action $a$ that yields a maximum value within the sum. The set $A(s)$ denotes all possible actions that can be taken in state $s$.

While Equation (\ref{bellman}) is a general representation applicable to stochastic MDPs, deterministic MDPs simplify the transition probability distribution $P(s' | s, a)$ to unity, signifying that there is only one possible next state for a given action and state. This mathematical foundation lays the groundwork for various RL algorithms, enabling agents to learn optimal policies and make informed decisions in complex and uncertain environments. Through the elegant interplay of mathematical principles and computational strategies, RL facilitates the development of intelligent agents capable of solving diverse real-world problems. In the field of quantitative finance, algorithmic trading involves making real-time decisions about where to trade, at what price, and how much to trade in a volatile and intricate financial market. Using various economic data, an RL trading system creates a multifaceted model for automated trading. This can be challenging for human traders to replicate. As a result, RL is seen as having a potential advantage as a tool in quantitative finance.

The unique flexibility of RL algorithms lies in their adaptability. They can be tailored to include multiple constraints or objectives. For example, an algorithm could be trained to maximize returns while also minimizing risk, or incorporating specific ESG criteria into investment decisions. This level of customization opens up a new opportunity in portfolio management, where sustainability and profitability go hand in hand. This is the feature that we exploit in this paper.

However, the current literature on the convergence of RL and ESG in portfolio management is still relatively sparse. While some studies have begun to incorporate ESG ratings using deep learning algorithms, they often focus on a narrow set of companies and rely on ESG scores from a singular source.  For instance, \cite{vo2019deep} firstly leveraged deep learning and incorporated ESG ratings into portfolio optimization by considering only six companies. \cite{maree2022balancing} not only included financial returns and risk (via a Sharpe ratio performance metric), and sustainability (via an ESG factor consideration) into their reward function, but they also experimented with several state-of-the-art RL agents, such as multi-agent deep deterministic policy gradients (MADDPG). This approach is problematic due to the noted inconsistencies in ESG ratings across different agencies. Basing an RL model on one source could lead to suboptimal investment decisions, given the discrepancies and potential biases in ratings. In addition, the recent work in \cite{garrido2023drl} explored how ESG considerations can be embedded into deep reinforcement learning (DRL) portfolio agents through reward design. Specifically, they simulate a regulated market environment where ESG-based financial incentives—taxes and grants—are imposed based on a portfolio’s ESG score relative to a benchmark index. This reward modification allows the agent to internalize ESG compliance incentives during policy learning. Their approach models ESG scores as fixed, externally validated signals and focuses on using regulatory mechanisms to influence agent behavior.

In contrast, our study addresses a more foundational and data-centric challenge: \textit{how to construct ESG-integrated portfolios in the presence of rating disagreement.} ESG ratings are known to be heterogeneous across providers (Berg et al., 2022), and investors face ambiguity not just in how ESG is valued, but in what ESG ratings to trust. Our contribution is thus complementary to policy-based approaches: where \cite{garrido2023drl} regulate how ESG matters, we address what to believe about ESG, and how those beliefs shape long-run portfolio performance under signal uncertainty.

Therefore, in this paper, we propose novel methods that can effectively incorporate ESG factors into investment decisions in the presence of widespread confusion among the investors with respect to the ESG ratings. Such research would be instrumental in ensuring that investment decisions align well with principles of sustainable finance and promote long-term sustainability in the economy.

\subsubsection{Overview of FinRL framework}

Creating an RL trading strategy can be challenging. The programming part is prone to errors and involves meticulous debugging. The development process includes tasks such as preparing market data, establishing a training environment, handling trading states, and evaluating trading performance through backtesting. In our empirical research, we employ a FinRL framework, introduced by \cite{liu2020finrl}, to address the complexities of financial decision-making within dynamic market environments. FinRL stands as a robust and adaptable platform specifically designed for applying reinforcement learning techniques to financial scenarios. The framework is uniquely tailored to cater to the intricacies of financial time-series data, offering a versatile foundation for developing and testing trading strategies, portfolio management techniques, and risk mitigation approaches. 

The FinRL package provides built-in functionalities to facilitate feature engineering, ensuring that the state variables in the RL framework are appropriately designed. These state features include OHLC prices, trading volumes, and various technical indicators such as Moving Average Convergence Divergence (MACD) and Relative Strength Index (RSI). These features are engineered to summarize historical information effectively, encapsulating dependencies in the current state. This approach supports the Markov property by ensuring that the state representation contains all necessary information for future decision-making, independent of earlier states. The dynamic interactions between the agent and the environment are managed within the \texttt{PortfolioEnv} environment provided by the FinRL framework. The environment is designed to handle state transitions based on the Markov assumption, where the next state depends only on the immediately previous state and the chosen action. By defining the state variables to include both the technical indicators and the portfolio-specific details (e.g., the account balance and the stock holdings), the environment ensures that the state transitions remain Markovian. This design simplifies the reinforcement learning process while maintaining the fidelity of the financial decision-making process.


The FinRL framework is structured as a three-layer system:
\begin{enumerate}
    \item \textbf{Environment Layer}: At the base, this layer emulates real financial markets using historical data. It includes essential metrics such as closing prices, stock quantities, trading volume, and technical indicators, creating a realistic environment for the agent's interaction.
    \item \textbf{Agent Layer}: This middle layer consists of RL algorithms that drive the agent's actions. Reward functions within this layer guide the agent's behavior as it navigates the market environment, enabling learning and decision-making through defined state and action spaces.
    \item \textbf{Application Layer}: At the top level, this layer represents practical applications of RL, including stock trading, portfolio allocation, and cryptocurrency trading. Here, the trading strategy is mapped into the RL framework by specifying key components like the state space, action space, and reward function.
\end{enumerate}

In the context of our research in this paper, we detail each layer as follows:

On the \textbf{application layer}, we map an algorithmic trading strategy into the RL framework by specifying the state space, action space, and reward function. 

\textbf{State Space:} The state space describes how the agent perceives the environment. A trading agent observes many features to make sequential decisions in an interactive market environment. In our calibration exercise, the states are set as follows:
\begin{itemize}
    \item Balance $b_t \in \mathbb{R}_{+}$: account balance at the current time step $t$. 
    \item Shares $\boldsymbol{k}_t \in \mathbb{Z}_{+}^n$ : current shares for each asset, where $n$ represents the number of stocks in the portfolio.
    \item Open-high-low-close (OHLC) prices $\boldsymbol{o}_t, \boldsymbol{h}_t, \boldsymbol{l}_t, \boldsymbol{p}_t \in \mathbb{R}_{+}^n$.
    \item Trading volume $\boldsymbol{v}_t \in \mathbb{R}_{+}^n$.
    \item Technical indicators: including Moving Average Convergence Divergence (MACD), and Relative Strength Index (RSI), which are momentum indicators widely used in momentum trading.
\end{itemize}

\textbf{Action Space:} The feasible set for actions in this RL framework consists of adjustments to portfolio holdings for each asset at each timestep. Specifically, the action space is defined as $a_t \in[-k, k]^n$, where $k$ represents a maximum allowable trading proportion (e.g., as a fraction of the current holding or the account balance) and $n$ is the number of assets in the portfolio. Actions are constrained by practical factors such as transaction costs, leverage limits, and liquidity considerations, ensuring that the trading strategy remains realistic and implementable. For example:
\begin{itemize}
    \item $a_t=-1$ : Sell all units of an asset.
    \item $a_t=0$ : Hold the current position.
    \item $a_t=1$ : Buy the maximum allowable units of an asset.
\end{itemize}

For instance, `Buy 10 shares of AAPL' or `Sell 10 shares of AAPL' correspond to actions of 10 or -10, respectively. 

\textbf{Reward Function:} We update the reward function by adding a weighted ESG score of the portfolio in the RL framework. It is assumed that the managing agent's reward is a linearly weighted function of the expected return and the mean ESG score for the portfolio.

\begin{equation}
\text{Reward}=\sum_{i=1}^n \omega_i r_{i, t}+\alpha \sum_{i=1}^n \omega_i \text{ESG}_i,
\end{equation} 
\noindent where $r_{i,t}$ is a standardized return of the $i^{th}$ stock at time $t$, $\text{ESG}_i$ is a standardized ESG score of the $i^{th}$ stock, and $\sum_{i=1}^n \omega_i=1, \omega_i \in[0,1]$. The parameter $\alpha$ determines the relative importance of financial returns versus ESG considerations in the reward function. In this model, $\alpha$ is set to 1 under the assumption that both the returns and the ESG scores are standardized, ensuring that they are on comparable scales. The returns are standardized by subtracting the mean and dividing by the standard deviation of the historical returns, while the ESG scores are transformed and standardized across their respective ranges (e.g., 0-100 or categorical ratings converted to numerical equivalents). This ensures both components contribute equally, facilitating a balanced optimization process of financial and sustainability objectives. The choice of $\alpha=1$ reflects an equal weighting, making it a suitable baseline for most investor profiles. Variations in $\alpha$ (e.g., higher for the ESG-focused investors or lower for the return-maximization) yield alternative portfolio preferences, suggesting its flexibility for tailoring the strategies.

Also, this reward function can change the form according to the different objective requirements, such as changing the reward to the Mean-Varianfe framework. This reward function is adaptable and can be reconfigured to accommodate different objective requirements. For example, it could be modified to align with a Mean-Variance framework, allowing the agent to optimize for both return and risk alongside ESG criteria.

On the \textbf{Agent layer}, FinRL allows users to plug in and play with the standard RL algorithm. In particular, we select a Deep Deterministic Policy Gradient (DDPG) algorithm as the agent as the main agent for our FinRL-based studies.  DDPG is a notable advancement in RL that operates within a continuous action space. It seeks to optimize the performance of deterministic policies by maximizing the objective function $J(\theta) = E_{s \sim p^\pi, a \sim \pi(\theta)}[R]$, where $p^\pi$ denotes the distribution of states and $\pi(\theta)$ represents the policy dictated by the parameter set $\theta$. DDPG operates by modeling the action selection process ($\mu_\theta: S \mapsto A$) through an actor-network, allowing it to predict an optimal action based on the current state. Concurrently, the algorithm approximates the reward function, represented by $Q^\mu(s, a)$, using a critic network. The critic network in DDPG is a neural network that estimates the value of a stateaction pair $(s, a)$ by predicting the expected cumulative reward when action $a$ is taken in state $s$ and the policy $\mu$ is followed thereafter. This score allows the agent to evaluate the quality of actions and iteratively refine its policy for improved decision-making over time.

A key insight within DDPG lies in the formalization of the gradient of the objective function for deterministic policies, which can be represented as:

\begin{equation}
\label{ddpg}
\nabla_\theta J(\theta)=E_s\left[\left.\nabla_\theta \mu_\theta(a \mid s) \nabla_a Q^\mu(s, a)\right|_{a=\mu_\theta(s)}\right]
\end{equation}

\noindent where the actor-network, denoted as $\mu_\theta: S \mapsto A$, serves as a function approximator guiding action selection based on the current state $s$, producing an anticipated action $a$ as per the policy parameter $\theta$. This mapping allows the agent to translate states into corresponding actions. The critic network, represented by $Q^\mu(s, a)$, approximates the value of state-action pairs $(s, a)$ under the policy $\mu$ by estimating the cumulative reward when taking action $a$ in state $s$ and subsequently following policy $\mu$. The gradient of the objective function, $\nabla_\theta J(\theta)$, indicates how changes in the policy parameter $\theta$ influence the goal of maximizing expected cumulative rewards. The expectation operator $E_s$ accounts for the average value over states, as dictated by the current policy $\mu_\theta$. The gradient $\nabla_\theta \mu_\theta(a \mid s)$ represents the impact of policy parameter adjustments on the actions forecasted by the actor-network for a given state $s$. Similarly, $\nabla_a Q^\mu(s, a)$ characterizes the effect of action changes on the estimated state-action value provided by the critic network. The expression $\left.\nabla_\theta \mu_\theta(a \mid s) \nabla_a Q^\mu(s, a)\right|{a=\mu_\theta(s)}$ captures the combined influence of policy parameter and action modifications on the overall objective. 
This equation outlines the mechanism for adjusting the policy's parameters to maximize cumulative rewards. It involves the computation of the gradient of the actor network's output with respect to the policy's parameters, multiplied by the gradient of the critic network's output with respect to the action, evaluated at the action predicted by the actor-network.

In the context of financial applications, integrating DDPG within the FinRL framework empowers us to train intelligent agents capable of making well-informed trading decisions, optimizing portfolio allocations, and navigating the intricacies of financial markets while considering continuous action spaces and the underlying uncertainties.

The \textbf{Environment layer} in FinRL is responsible for observing current market information and translating that information into states of the MDP problem. The state variables can be categorized into the state of an agent and the state of the market. In our study, the state of the market includes the open-high-low-close prices and volume (OHLCV) and technical indicators (MACD and RSI); the state of an agent includes the account balance and the shares for each stock.

The following is an algorithmic representation (See Algorithm \ref{algorithm}) that provides a detailed breakdown of the steps involved in RL with the Deep Deterministic Policy Gradient (DDPG) agent.

\begin{algorithm}
\caption{RL with DDPG Agent for Portfolio Optimization}
\begin{small} 
\label{algorithm}
\begin{algorithmic}[1]
\STATE \textbf{Input:}
\STATE \quad State variables: Portfolio data (holdings, cash balance), market data (OHLC prices, technical indicators).
\STATE \quad Hyperparameters: Discount factor $\gamma$, learning rates, noise parameters.
\STATE \quad Custom-designed reward function (e.g., portfolio returns and metrics like ESG scores).
\STATE \quad Replay buffer size, batch size, and number of episodes.
\STATE
\STATE \textbf{Initialization:}
\STATE \quad Initialize actor network $\mu(s|\theta^\mu)$ and critic network $Q(s, a|\theta^Q)$ with random weights.
\STATE \quad Initialize target networks $\mu'$ and $Q'$ as copies of the actor and critic networks.
\STATE \quad Initialize replay buffer $\mathcal{D}$ to store experience tuples $(s_t, a_t, r_t, s_{t+1}, \text{done})$.
\STATE
\STATE \textbf{Procedure:}
\FOR{each episode}
    \STATE Reset the environment and initialize $s_0$.
    \FOR{each timestep}
        \STATE Select action:$a_t = \mu(s_t|\theta^\mu) + \text{noise}$
        \STATE Execute $a_t$ in the environment.
        \STATE Observe reward $r_t$, next state $s_{t+1}$, and done flag.
        \STATE Store transition $(s_t, a_t, r_t, s_{t+1}, \text{done})$ in $\mathcal{D}$.
        \STATE Sample mini-batch of $N$ transitions from $\mathcal{D}$ and update:
        \STATE \quad \textbf{Critic:}
        $$y_i = r_i + \gamma Q'(s_{i+1}, \mu'(s_{i+1}|\theta^{\mu'}))$$
        Minimize Bellman error:
        $L = \frac{1}{N} \sum_i (y_i - Q(s_i, a_i|\theta^Q))^2$
        \STATE \quad \textbf{Actor:}
        Update using policy gradient:
        $$\nabla_{\theta^\mu} J \approx \frac{1}{N} \sum_i \nabla_a Q(s, a|\theta^Q)|_{s=s_i, a=\mu(s_i)} \nabla_{\theta^\mu} \mu(s|\theta^\mu)|_{s_i}$$
        \STATE Periodically update target networks:
        $$\theta^{Q'} \leftarrow \tau \theta^Q + (1 - \tau) \theta^{Q'}$$
        $$\theta^{\mu'} \leftarrow \tau \theta^\mu + (1 - \tau) \theta^{\mu'}$$
        where $\tau$ is the target update rate (e.g., $\tau \in [0.001, 0.01]$).
    \ENDFOR
\ENDFOR
\STATE
\STATE \textbf{Output:} Trained actor network $\mu(s|\theta^\mu)$ for optimal portfolio allocation.
\end{algorithmic}
\end{small} 
\end{algorithm}

\newpage

The process of training through RL consists of monitoring price fluctuations, making decisions (selling, holding, or purchasing a specific quantity of stock), and computing a reward. By interacting with the environment, the agent updates iteratively and eventually obtains a trading strategy to maximize the expected return. ESG ratings are sourced from four major providers: RobecoSAM, Sustainalytics, MSCI, and Refinitiv Asset4. These ratings cover large-cap U.S. equities and are obtained via subscription to Wharton Research Data Services (WRDS). Financial data and relative features used in the model are also sourced from CRSP and Compustat via WRDS, and matched with ESG ratings based on firm identifiers and reporting periods. To calibrate the portfolio selection process by using an RL framework, we download daily data for DOW 30 from 06/30/2007 to 06/30/2022. This time window is chosen to ensure consistent ESG rating availability across all four major providers as ESG coverage becomes more comprehensive from 2007 onward. We adopt a rolling-window evaluation framework commonly used in time-series analysis to simulate realistic out-of-sample investment decisions. Each rolling window consists of a 3-year training period followed by a 1-year testing period. The first window covers 06/30/2007 to 06/30/2011, with the first three years used for model training and the final year for testing. Subsequent windows follow the same format, resulting in a total of 12 distinct evaluation periods, labeled Period 1 through Period 12. This setup allows the model to capture temporal variation in both ESG signal quality and market dynamics, while balancing the need for sufficient training data with the risk of overfitting to outdated information. Within each window, we train and evaluate RL agents using ESG scores from different rating providers to analyze strategy performance under varying signal conditions.

For concreteness, we introduce a general flow chart of RL under the context of our study in Figure \ref{fig1}. This approach compares several key components:

\begin{figure}[!htp]
\includegraphics[width=12cm]{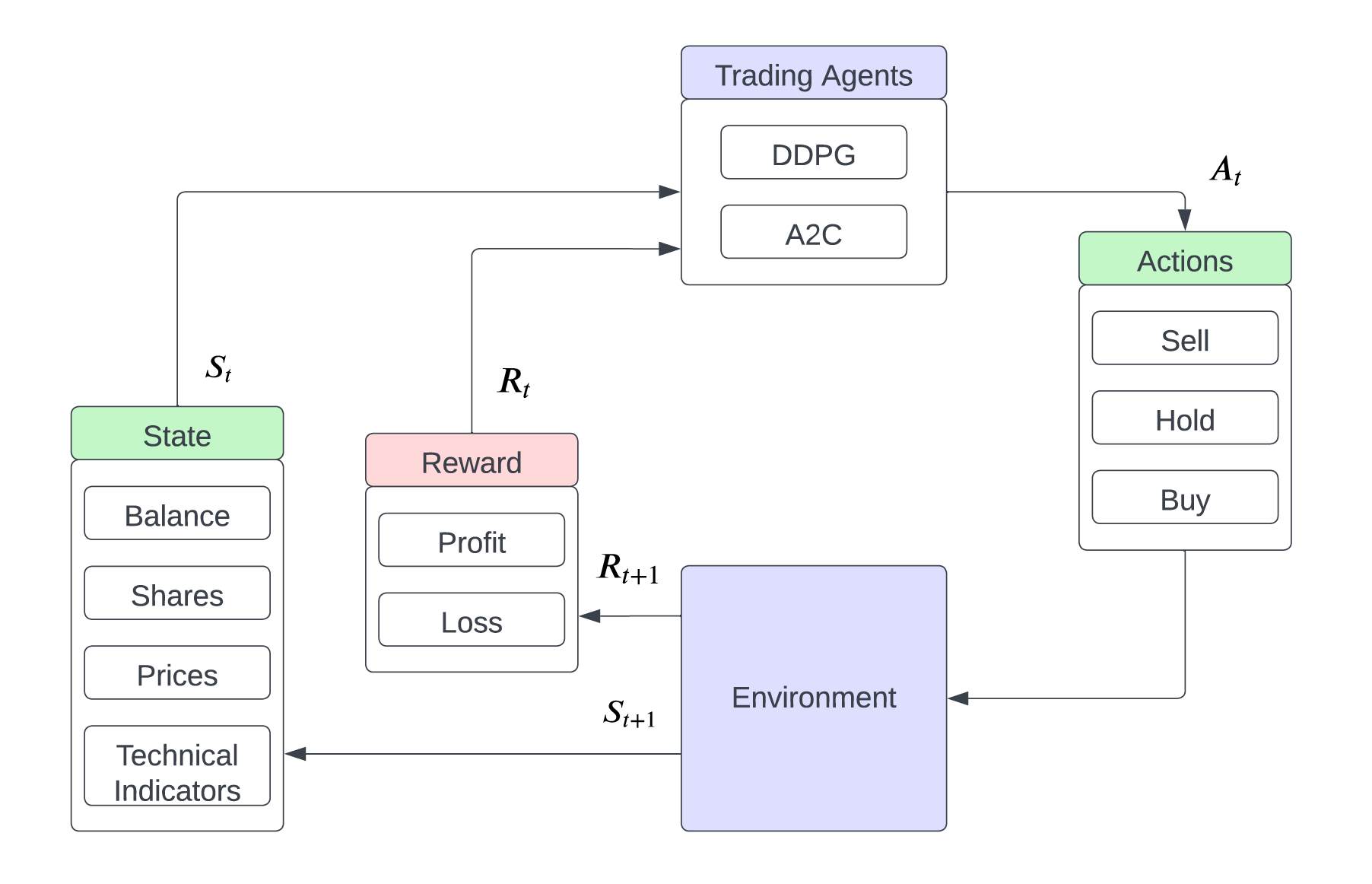}
\centering
\caption{A general flow chart of FinRL applied in portfolio selection}
\label{fig1}
\end{figure}

\begin{figure}[H]
\includegraphics[width=14cm]{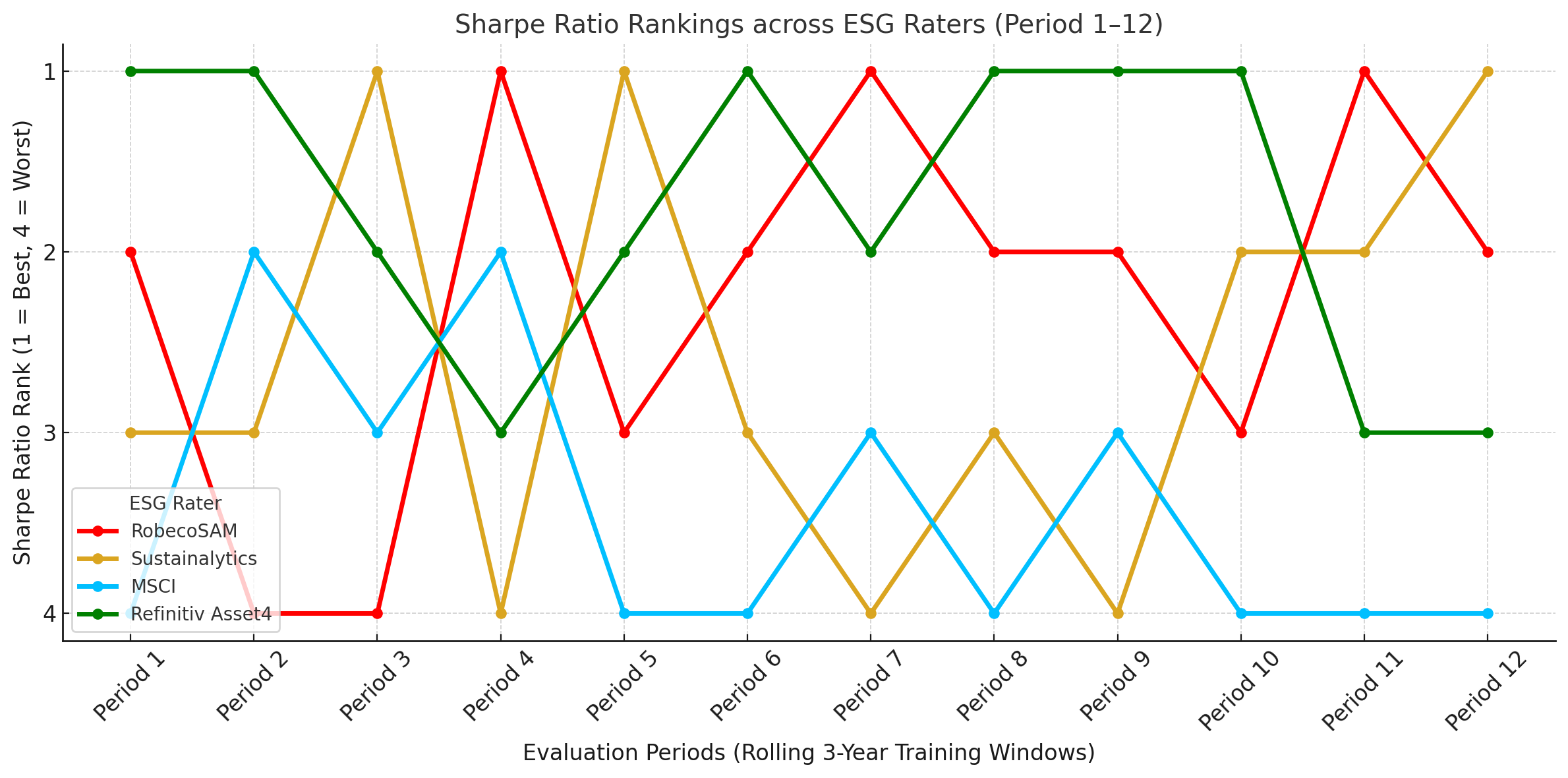}
\centering
\caption{Sharpe Ratio Rankings across ESG Raters over Time}
\label{fig2}
\end{figure}

Each period involves establishing the RL environment, determining states and actions, and modifying reward functions as previously described. Specifically, the reward function is updated through a linear combination of the portfolio return and the weighted ESG score from each ESG rater for the corresponding period. We compare the values of the reward function, defined as a weighted combination of the portfolio market returns and the ESG scores, by varying the ESG ratings sourced from different raters over the same time period. The rankings of the reward function for the four raters across the 12 periods from 2011 to 2022 are visually depicted in Figure \ref{fig2}. Each line represents the relative Sharpe ratio ranking (1 = best, 4 = worst) of portfolios constructed using ESG scores from a given rater across twelve rolling time periods. The code is implemented in Python by utilizing the open-source FinRL framework created by \cite{liu2020finrl}, which provides a well-calibrated set of hyperparameters optimized for financial market applications. As a primary objective of this research is to evaluate the incorporation of ESG factors into portfolio management, rather than to optimize the RL algorithm performance, no additional fine-tuning of hyperparameters is performed. The FinRL framework includes pre-tuned settings that have been tested extensively on various financial markets, ensuring robustness and efficiency for standard scenarios. These pre-configured parameters, along with the rolling-window approach used in training, are sufficient for the goals of this study.

The figure shows that there is no single rater that is capable of producing a portfolio that performs the best consistently over all of the 12 combinations of moving windows under consideration. Instead, we observe that the rankings appear to switch back and forth among the four raters for the time periods under study. This vividly illustrates the impacts of a widespread confusion among investors who are uncertain about which ESG rater to rely on to identify the best-performing portfolios from the ranked companies. Additionally, the divergence in ESG ratings has a real potential to undermine the confidence of companies seeking to improve their ESG performance.

\section{Addressing the heterogeneity of ESG raters}
\label{2.3}

\subsection{Data-driven ensemble strategies}

ESG inconsistency poses substantial challenges, often stemming from issues related to data availability, quality, and bias. A key problem for this inconsistency is the lack of sufficient or accurate ESG metrics, as many datasets are incomplete or fail to cover the full spectrum of environmental, social, and governance factors. The quality of available data is another concern, with discrepancies arising from differences in reporting standards and methodologies across regions. Additionally, some ESG data may be published to meet specific objectives, potentially introducing bias in favor of certain agendas or stakeholders. For instance, ESG frameworks were introduced predominantly in developed countries, which may not adequately represent the challenges and priorities of emerging economies. These limitations can lead to inconsistent evaluations, and potentially reduce the reliability of ESG assessments.

In investigating factors for the reported divergence of ESG ratings, \cite{berg2022aggregate} decompose the ratings into three sources: (1) $56 \%$ of the ranking heterogeneity can be traced back to purely measurement issues; (2) $38 \%$ of the ranking disagreement is attributed to scope, where different raters include different attributes; and (3) $6 \%$ of the rating differences is due to weight placed by different raters on individual components (E, S, and G) when they calculate their aggregate ESG scores for the companies. Furthermore, \cite{berg2022aggregate} point out that there is even a divergence of ratings on verifiable facts from public records. Additionally, there exists a so-called `rater effect', where a company rated highly on a particular attribute enjoys high ratings on its other attributes.

The question of interest to us is: {\it what are the main implications of high heterogeneity in ESG rankings of companies to investors in terms of their asset allocation?} It certainly does not imply that ESG ratings are so unreliable that they should not be used by investors as factors in forming their investment strategies. After all, a company's performance in general is known to be difficult to measure with any degree of accuracy. However, it does mean that investors need to acquire a better understanding of what the ESG ratings published by different rating agencies are measuring.

Ambiguity (also known as Knightian uncertainty in microeconomics literature) and risk are two related but distinct concepts in Decision Theory. Ambiguity refers to uncertainty (or lack of clarity) with unknown odds, while risk refers to the potential for loss or failure with known odds. Investors can have multiple views when forming their investment strategies. In general, investors' views may vary in their level of optimism (or conservatism); as a result, these views may be associated with more or less favorable scenarios. However, it is impossible to know with certainty which views are more likely to be relatively more accurate, so investors must weigh the pros and cons of each view in order to arrive at their final decision on investment. In this paper, ambiguity specifically refers to the uncertainty surrounding these different views. 

In this paper, investors are regarded to have incomplete information about ESG scores from different raters. They are not sure which one is right and what are the better weights of different ESG scores. In particular, each ESG rater can be regarded by investors as a `view'. Hence, confusion about ESG ratings can be regarded as a type of ambiguity. In fact, each rater has its own evaluation system. It is hard for investors to figure out which ESG rater provides a more accurate view of the ESG score and it is also not easy for investors to determine the most effective way to combine or incorporate these different views. 

Instead of attempting to identify an optimal universal ESG representative for any given time period, we take an agnostic approach in this paper. We provide several strategies to ensemble the ESG scores based on investors' attitudes toward the ambiguity of ESG ratings. This involves considering methods that essentially form certain weights among different ESG raters. The assemble strategies that we propose in this paper are:
    
    \textbf{Clustering centroid}: The centroid of a group of points is a point that represents the center or average of the group. For example, in a data analysis context, the centroid of a cluster of data points might be used as a reference point to calculate the dissimilarity of each point to the centroid. For each company, we use Euclidean distance to measure a `dissimilarity' between the ESG rater and the centroid. The centroid measures a minimum sum of the dissimilarity of the values in a dataset. After a trivial derivation, the centroid will reduce to the sample mean of the four raters. The centroid gives us an idea of where the `center value' is located among the raters and this is an appropriate criterion if the investor takes all observations into consideration without losing any information. In other words, the centroid represents a decision-maker who places the same weight on those different views. However, the simple mean is known to be misleading when the data sample is skewed or contains outliers.

    \textbf{Median}: The median is a measure of central tendency that is resistant to outliers or extreme values. Ambiguity-averse investors may prefer to use the median as their decision-making tool because it is a more stable and reliable measure of central tendency that is less influenced by extreme views or opinions. This can help them deal with ambiguity and make more informed investment decisions. Thus, if the ratings of the investor's target company differ greatly among the rating agencies, the median is a more reliable metric in terms of its robustness.

    \textbf{Principal component analysis (PCA)}: PCA is a popular data-driven technique that is used to reduce the dimensionality of a dataset. It does this by transforming the data into a new coordinate system, in which the dimensions or variables are made to be orthogonal. The new dimensions are called principal components, and they are ordered by the amount of variation or information that they capture from the original data.
    
    PCA is often used to decide on a representative from a set of choices because it can help to identify the underlying structure or patterns in the data. By reducing the dimensionality of the dataset, PCA can reveal the most important variables or dimensions that capture the majority of the variation in the data. This can be useful for identifying the key factors or characteristics that are driving the variation in the data, and for selecting a representative that captures the essence of the dataset as a whole.

    \textbf{Alpha-Maxmin}: \cite{GMM04} characterize the ambiguity attitude by a linear combination of the worst view and the best view, which they term as an $\alpha$-\textit{maxmin expected utility } ($\alpha$-MEU) form,
    \begin{equation}
    \label{alpha-maxmin}
    \alpha \inf _{\mathbb{Q} \in \mathcal{Q}} \mathbb{E}^{\mathbb{Q}}[U(X)]+(1-\alpha) \sup _{\mathbb{Q} \in \mathcal{Q}} \mathbb{E}^{\mathbb{Q}}[U(X)],
    \end{equation} 
    where $U$ represents a Bernoulli utility function that captures the decision-maker's preferences over outcomes, typically reflecting risk attitudes such as risk aversion (e.g., $U(X)=\log (X)$ for a logarithmic utility or $U(X)=X^{1-\gamma} /(1-\gamma)$ for a constant relative risk aversion, where $\gamma$ denotes a risk aversion parameter). The variable $X$ denotes uncertain outcomes or payoffs, such as returns or wealth generated by a portfolio of assets, with its definition depending on the context (e.g., future cash flows or asset returns). The set $\mathcal{Q}$ is a collection of subjective probability measures or priors that represent the decision-maker's ambiguity about the true distribution of $X$, capturing a Knightian uncertainty (or ambiguity). This framework generalizes the $\alpha$-maxmin rule of Hurwicz to settings of Knightian uncertainty, where a subjective perception of ambiguity is represented by $\mathcal{Q}$ and the parameter $\alpha \in[0,1]$ captures the attitude toward ambiguity, balancing pessimism (worst-case scenarios) and optimism (best-case scenarios).
    
    Unlike the first three metrics, which are purely data-driven, the Alpha-Maxmin metric combines a behavioral criterion on the part of the investors with a data-driven criterion. Also, unlike the median, Alpha-Maxmin takes into account the most extreme cases among investors' ambiguous views. This means that Alpha-Maxmin considers not just the average or typical view or opinion of investors, but also the most optimistic and most pessimistic perspectives of investors. Therefore, investors who are more concerned about the likelihood of extreme outcomes may prefer to use Alpha-Mammin as their decision-making tool.

These four metrics will be used to form ensemble strategies by combining different investment styles to create a well-rounded portfolio that meets the investor's goals. This is acheived by incorporating investors' ambiguous views of ESG ratings in the RL framework given the prevalent heterogeneity of the ratings.

\subsection{Steps to Improve ESG Measure Consistency}

The high degree of inconsistencies in ESG measures provided by rating agencies like Bloomberg, Thomson Reuters, and Standard \& Poor's can be attributed to their differences in scope, measurement methodologies, and the weight assigned by them to individual ESG factors. To address these issues and enhance the consistency of the ESG measures from the sources, several steps are possible:
\begin{enumerate}
    \item Standardization of Metrics: Establishing a universally agreed-upon framework for ESG metrics is essential. Agencies should collaborate to define key indicators under Environmental, Social, and Governance categories. For example, standardized environmental metrics could include carbon emissions intensity, while social metrics could cover gender diversity rates. This would reduce variability caused by differing scopes across agencies.
    \item Transparent Methodologies: ESG rating agencies should disclose their methodologies in greater detail, including how they collect data, weigh individual metrics, and calculate overall ESG scores. Increased transparency would allow investors to better understand and compare ratings across agencies.
    \item Third-party Oversight: An independent body or consortium could be established to oversee and audit ESG rating methodologies. This body could provide certifications for agencies that adhere to standardized processes, similar to how financial auditing works under frameworks such as International Financial Reporting Standards (IFRS) or Generally Accepted Accounting Principles (GAAP).
    \item Data Integration: Agencies could adopt ensemble techniques, as proposed in this study, to reconcile the divergent ESG scores. These techniques use a combination of scores from multiple raters to create a more balanced and robust ESG rating, mitigating the impact of individual agency biases or methodological differences.
\end{enumerate}
By implementing these steps, the financial industry can move closer to achieving greater consistency in ESG measures, fostering trust and facilitating informed decision-making by investors. Our findings further emphasize the need for such standardization, as the current inconsistencies not only create confusion but also limit the scalability of ESG investing.

\section{ESG-related portfolio with mean-variance preference} 
\label{2.4}

In the previous section, we have treated the heterogeneity in ESG scores generated by different ESG raters as a form of ambiguity within the realm of sustainable investments. In this section, our initial focus is on a single category of ESG scores in the presence of a general notion of uncertainty, in order to allow us to explore the linkages between risk and uncertainty which was emphasized in recent studies such as \cite{pastor2021sustainable} and \cite{avramov2022sustainable}.  

We will return to the concept of ambiguity in the empirical analysis later on. Our current focus in this section is on unraveling distinct analytical ways in which the identical ESG score type influences portfolio performance through the lens of uncertainty in general. This approach underscores our commitment to providing a nuanced and in-depth understanding of the intricate dynamics at play within the realm of sustainable investments. We assume a linear mean-variance (MV) preference structure in the Double Mean-Variance (DMV) model to ensure analytical tractability and clear separation of pecuniary and non-pecuniary returns. This linearity assumption facilitates closed-form solutions for optimal portfolio weights and enables transparent interpretation of risk and ESG-related uncertainty. While a linear preference simplifies the analysis, it may overlook potential nonlinear behaviors in actual investor utility functions. Future research could extend this framework to accommodate non-linear preferences or prospect theory-based utilities.

To frame our discussion, we start with a general analytical framework initially. Instead of formulating investors' preferences with an exponential utility (CARA) function as in \cite{pastor2021sustainable} and \cite{avramov2022sustainable}, we use a mean-variance (MV) preference to formulate the utility function. Our analysis draws heavily from \cite{maccheroni2013alpha}, 
which shows that the optimal investment strategy under the MV preference has an analytical closed-form expression, which provides a sharp separation for key factors of interest, such as expected return and risk. Most importantly, \cite{maccheroni2013alpha} show that for a
von Neumann-Morgenstern expected utility maximizing investor with a utility function $u(\cdot)$ representing 
investor's attitude toward risk, and wealth $a$, who considers an investment project $h$, a linear Arrow-Pratt approximation of her certainty equivalent for an uncertain investment prospect is  given by:
\begin{equation}
c\left(a + h, P \right) \approx
a + \mathbb{E}_P(h) - \frac{1}{2} \gamma_u (a) \sigma_P^2(h)
\label{ch2.25}
\end{equation}
where $c(\cdot)$ is a first-order certainty-equivalent measure and $\gamma_u(a)=-u^{\prime \prime}(a) / u^{\prime}(a)$ is a parameter that links risk premium and volatility. In Equation (\ref{ch2.25}), both $\mathbb{E}_P (\cdot)$ and $Var_P(\cdot)$ are taken with respect to a particular probabilistic model (denoted by $P$), which describes a stochastic nature of the investment problem. 

This approximation leads to an MV model with a prospect $f$ 
being evaluated at:
\begin{equation}
U(f) = \mathbb{E}(f) - \frac{\gamma_u(a)}{2} 
\sigma^2(f)
\end{equation}
with $f=a+h$ and $\gamma = \gamma_u(a)$.

MV analysis is a crucial preliminary step in our investment strategy due to its direct relevance to the subsequent implementation of the Capital Asset Pricing Model (CAPM). By employing MV analysis, we establish a foundational understanding of risk-return trade-offs that informs the CAPM application. MV analysis allows us to evaluate the relationship between an asset's expected returns and its associated risk, thus enabling the identification of an efficient frontier of portfolios that offer the highest possible return for a given level of risk or the lowest possible risk for a targeted level of return. This efficient frontier forms the foundation for CAPM's principles, as it provides a range of diversified portfolios that theoretically maximize returns based on an investor's risk tolerance. Therefore, the judicious application of MV analysis not only refines the portfolio construction but also paves the way for an extension of the Modern Portfolio Theory (MPT), proposed by \cite{Markowitz}, through the subsequent utilization of CAPM's insights.

\subsection{Model setting}

With additional efforts, it is possible to derive our ensuing analytical results directly from the general framework set up above. However, we do not pursue this strategy in this paper. Instead, we present a model of an ESG-related portfolio with MV preferences in the presence of general uncertainty in the sense of \cite{pastor2021sustainable} and \cite{avramov2022sustainable}.  
More specifically, in a single-period economy, suppose that there is an agent who trades at time 0 and closes her position at time 1. Denote a random rate of return on the market portfolio as $\tilde{r}_M$, and a true and unobservable ESG score of the market portfolio as $\tilde{g}_M$. Let $\mu_f$ be a risk-free rate. We model the excess returns of the market and the ESG factor as follows:
\begin{equation}
Market: \quad \tilde{r}_M-\mu_f=\mu_M+\tilde{\epsilon}_M, 
\end{equation}
and
\begin{equation}
ESG: \quad \tilde{g}_M=\mu_{g}+\tilde{\varepsilon}_{g, M},
\end{equation}
where $\mathbb{E}[\tilde{r}_M]=\mu_M+\mu_f$ is an expected market return, $\mathbb{E}[\tilde{g}_M]=\mu_{g}$ is an expected value of the ESG-factor return, and $\tilde{\epsilon}_M$ and $\tilde{\epsilon}_{g, M}$ are zero-mean error terms. We also define the term $\sigma_M^2=\boldsymbol{X}_M^{T} \boldsymbol{\Sigma}_M \boldsymbol{X}_M$ and $\sigma_{g}^2=\boldsymbol{X}_M^{T} \boldsymbol{\Sigma}_{g,M} \boldsymbol{X}_M$ to respectively represent the variance of market returns and the variance of the ESG-factor returns respectively. Here, $\boldsymbol{X}_M$ represents a column vector of portfolio weights, and $\boldsymbol{\Sigma}_M$ and $\boldsymbol{\Sigma}_{g,M}$ represent the covariance matrix of market returns and covariance matrix of ESG-factor returns respectively.

Next, denote $w$ as a weight of risky assets in a portfolio, i.e. the weight of the market portfolio. The parameters $\gamma$ and $\theta$ represent the investor's attitude toward risk associated with the market return and uncertainty associated with the ESG-factor return respectively. The objective functions for the market and the ESG factor are expressed respectively as:

\begin{equation}
    Market \quad Objective: w \cdot \mu_M+(1-w) \mu_f-\frac{\gamma}{2} w^2 \sigma_M^2,
\end{equation}
and
\begin{equation}
    ESG \quad Objective: w \cdot \mu_{g}-\frac{\theta}{2} w^2\sigma_{g}^2.
\end{equation}
The total expected utility function leads to a linear combination of the expected MV utility function of the market return and the expected MV utility function of the ESG-factor return, which gives rise to a new model that we dub as a Double MV (DMV) model:
\begin{equation}
    w \cdot \mu_M+(1-w) \mu_f-\frac{\gamma}{2}w^2 \sigma_M^2 +b\left(w \cdot \mu_{g}-\frac{\theta}{2} w^2\sigma_{g}^2 \right),
\end{equation}
where $b > 0$ captures the relative taste of the investor between `pecuniary' return and `non-pecuniary' return. 

Pecuniary return refers to a financial gain or loss from an investment, such as the change in the value of the investment or the income generated by the investment, such as dividends or interest. Pecuniary return is typically measured in monetary units, such as Dollars or Euros. Pecuniary return is an important consideration for investors because it represents the financial gain or loss from the investment. Investors typically want to maximize their pecuniary return in order to achieve their financial goals, such as building up wealth or generating income.

Non-pecuniary return, on the other hand, refers to non-monetary benefits or costs of an investment project, such as social, environmental, or ethical impacts of the investment. Non-pecuniary return is often referred to as `externalities', as it represents the effects of the investment on parties that are not directly involved in the investment decision, which can also impact the overall performance and risk profile of the investment. For example, if a company has poor environmental practices, it may face regulatory risks or reputational damages that could negatively impact its financial performance. Similarly, if a company has poor working conditions or engages in unethical practices, it may face reputational risks or consumer backlashes that could also impact its financial performance.

Therefore, by considering both pecuniary and non-pecuniary returns in the model, we allow the investor to make informed investment decisions that take into account a full range of potential risks, uncertainties, and rewards of the investment universe. It can also help investors align their investments better with their values and ethical principles.
Next, we divide the value of $b$ into three cases:
\begin{itemize}
    \item If $b > 1$, the investors will place more emphasis on non-monetary gains.
    \item If $b < 1$, the investors will place more emphasis on monetary gains.
    \item If $b = 1$, the investors will be indifferent between monetary and non-monetary gains.
\end{itemize}

The optimal portfolio weights are obtained from the first necessary order condition of the underlying optimization program:
\begin{equation}
    \mu_{M}-\mu_{f}-\gamma w\sigma_M^2 +b\left(\mu_{g}-\theta w\sigma_{g}^2\right)=0
\end{equation}
as
\begin{equation}
    w=\frac{\mu_M-\mu_f+b \mu_{g}}{\gamma \sigma_M^2+b \theta {\sigma}_{g}^2}
\label{equweigt}
\end{equation}

This optimal portfolio weight from \ref{equweigt} can be rewritten as:
\begin{equation}
w =\frac{1}{\gamma} \frac{\left(\mu_M-\mu_f\right)}{\sigma_M^2}+\frac{1}{\gamma} b \frac{\mu_{g}}{\sigma_M^2} \\ 
-\frac{1}{\gamma} \frac{\mu_M-\mu_f+b \mu_{g}}{\sigma_M^2}\left(\frac{b \theta \sigma_{g}^2}{\gamma \sigma_M^2+b \theta \sigma_{g}^2}\right),
\label{equ221}
\end{equation}
where we can obtain three terms:
\begin{itemize}
    \item First Term $\frac{1}{\gamma} \frac{\left(\mu_M-\mu_f\right)}{\sigma_M^2}$: There is no parameter in this term which involves the ESG factor, and it serves as a benchmark case of ESG indifference.
    \item Second Term $\frac{1}{\gamma} b \frac{\mu_{g}}{\sigma_M^2}$: This term involves the return of the ESG factor, but it does not incorporate uncertainty with respect to the ESG rating. Thus, it serves as a benchmark case with ESG preferences when the ESG profile is known with certainty.
    \item Last Term $\frac{1}{\gamma} \frac{\mu_M-\mu_f+b \mu_{g}}{\sigma_M^2}\left(\frac{b \theta \sigma_{g}^2}{\gamma \sigma_M^2+b \theta \sigma_{g}^2}\right)$: This term captures the incremental effect of ESG uncertainty.
\end{itemize}

Next, we define three types of investors, which have ESG indifference (I), ESG preference with no uncertainty (N), and ESG preference with uncertainty (U). 
\begin{itemize}
    \item Type-I Investors: These investors do not consider ESG factors in their decision-making processes about their decisions. They are indifferent with respect to the impacts of their investments and focus solely on their profit considerations.
    \item Type-N Investors: These investors prioritize ESG factors and try to align their investments with their values. However, they do not face any uncertainty associated with the ESG scores. This implies that they solely factor in the expected portfolio ESG score and do not take the ESG score's variability into account.
    \item Type-U Investors: These investors also prioritize ESG factors, but they face uncertainty about the impacts of their investments on these factors. Consequently, these investors must grapple with the risk arising from variability in ESG scores. This complexity can render it more challenging for these investors to harmonize their investments with their values.
\end{itemize}

To derive a market premium, we set the first term in the above equation to 1 and derive the excess return of investors of type I. Similarly, letting the first two terms be 1 will lead to the formation of an excess return for investors of type N. Finally, the excess return for investors of type A is derived from the original weight form $\frac{\mu_m-\mu_f+b \mu_{g,m}}{\gamma \sigma_m^2+b \theta {\sigma}_{g}^2}=1$. After re-arranging terms in these formulas, the excess returns of the three types of investors can be expressed succinctly as:
\begin{itemize}
    \item Type-I Investors:
    \begin{equation}
        \mu_M^I-\mu_f=\gamma \sigma_M^2
    \end{equation}
    \item Type-N Investors: 
    \begin{equation}
        \mu_M^N-\mu_f=\gamma \sigma_M{ }^2-b \mu_{g}
    \end{equation}
    \item Type-U Investors: 
    \begin{equation}
        \mu_M^U-\mu_f=\gamma\sigma_M^2+b \theta \sigma_{g}^2-b \mu_{g}
    \end{equation}   
\end{itemize}

Hence, the differences in expected returns between these three types of investors are given respectively by:
\begin{equation}
    \mu_M^N-\mu_M^I=-b \mu_{g} <0
\end{equation}

\begin{equation}
    \mu_M^U-\mu_M^N=b \theta \sigma_{g}^2 >0
\end{equation}

\begin{equation}
    \mu_M^U-\mu_M^I=b \theta \sigma_{g}^2-b \mu_{g,M} = b(\theta \sigma_{g}^2 -\mu_{g})
\end{equation}

If we assume that $b>0$, then the no-uncertainty case is associated with a negative ESG incremental premium. \cite{pastor2021sustainable} pointed out that sustainable investing produces a positive social impact by making firms greener and by shifting real investment toward green firms. In order to achieve the gains from ESG, investors are willing to sacrifice their returns. Green assets have low expected returns because investors enjoy holding them and also because green assets hedge climate risk. 

In addition, it is evident from the second equation that the market premium increases with ESG uncertainty. The uncertainty of ESG scores contributes to additional risk for investors; so the required return should reflect the corresponding compensation for higher uncertainty. However, the sign of the incremental premium is indeterminate when we compare the ESG indifference case and the ESG preference with uncertainty case due to the conflicting forces emanating from the variance and the return of the ESG scores. This can be explained as follows: the investors who prefer non-monetary returns with ESG uncertainty will get compensation from the risks of the ESG uncertainty as well as take the loss from the non-monetary return. 

To derive the portfolio variances for the three types of investors (who are ESG indifferent, without ESG uncertainty, and with ESG uncertainty), we substitute the respective expressions for the weights $(w)$ into the general portfolio variance formula, $\sigma^2=w^2 \cdot \sigma_M^2$, where $w$ is the weight of the market portfolio and $\sigma_M^2$ is the variance of the market return. The weights for the three investor types are derived from the optimal portfolio weight decomposition in Equation (\ref{equ221}):
\begin{itemize}
    \item ESG Indifference ( $w^I$ ): The weight corresponds to only the first term of Equation (\ref{equ221}), $\frac{1}{\gamma} \frac{\mu_M-\mu_f}{\sigma_M^2}$. This reflects the benchmark case of ESG indifference, where the portfolio weight depends solely on the market return premium ( $\mu_M-\mu_f$ ) relative to the market risk, with no consideration of any ESG factors.
    \item Without ESG Uncertainty $\left(w^N\right)$: The weight includes the first two terms of Equation (\ref{equ221}):
    $$
    \frac{1}{\gamma} \frac{\mu_M-\mu_f}{\sigma_M^2}+\frac{1}{\gamma} \frac{b \mu_g}{\sigma_M^2} .
    $$
    This formulation adds a deterministic ESG premium $\left(b \mu_g\right)$ to the portfolio weight, reflecting an investor who accounts for the ESG factors but does not consider the uncertainty associated with them.
    \item With ESG Uncertainty $\left(w^U\right)$: The weight incorporates all three terms from Equation (\ref{equ221}), or equivalently, uses the full form of Equation (\ref{equweigt}):
    $$
    \frac{1}{\gamma} \frac{\mu_M-\mu_f+b \mu_g}{\sigma_M^2} \cdot\left[1-\frac{b \theta \sigma_g^2}{\gamma \sigma_M^2+b \theta \sigma_g^2}\right] .
    $$
    This weight adjusts for both ESG premium and its associated uncertainty. 
\end{itemize}
By substituting these weights into the variance formula, we obtain the portfolio variances for the three types of investors, reflecting their different attitudes toward ESG factors and uncertainty. Next, we derive the differences in the portfolio variances for each pair of the three investor types:

\begin{equation}
    \sigma_M^N-\sigma_M^I=\frac{2\left(\mu_M-\mu_f\right) b \mu_{g}+b^2 \mu_{g}^2}{\gamma^2 \sigma_M^2}>0,
\end{equation}
and
\begin{equation}
    \sigma_M^U-\sigma_M^N=-\frac{\left(\mu_M-\mu_f+b \mu_{g}\right)^2}{\gamma^2 \sigma_M{ }^2} \cdot \left[\frac{2 \gamma \sigma_M^2 b \theta \sigma_{g}{ }^2+b^2 \theta^2 \sigma_{g}^4}{\left(\gamma \sigma_M^2+b \theta \sigma_{g}^2\right)^2}\right] <0,
\end{equation}
and
\begin{equation}
    \sigma_M^U-\sigma_M^I=\frac{\left(\mu_M-\mu_f\right)^2\left[E_g-1\right]+\left[2\left(\mu_M-\mu_f\right) b \mu_{g}+b^2 \mu_{g}^2\right] E_g}{\gamma^2 \sigma_M^2},
\end{equation}
where $E_g = \frac{\left(\gamma \sigma_M^2\right)^2}{\left(\gamma \sigma_M^2+b \theta {\sigma}_{g}^2\right)^2} \in (0,1)$, which can be interpreted as a ratio of the risk of market returns to the overall risk after taking into account the degree of aversion. 

The variance relationships between different investors are generally the reverse of the return relationships, with the exception of the indeterminate relationship between the ESG indifference case and the ESG preference with uncertainty case. 


\begin{table}[h!]
\begin{tabular}{|l|l|l|l|}
\hline & \textbf{Return} & \textbf{Variance} & \textbf{Sharpe Ratio} \\
\hline
\hline
ESG Indifference & Large & Small & Large \\
\hline 
Without Uncertainty & Small & Large & Small \\
\hline 
With ESG Uncertainty & Around Large & Around Small & Around Large \\
\hline
\end{tabular}
\centering
\caption{Ex-ante Relationships between Different Types of Investors}
\label{tab231}
\end{table}

Thus, it is reasonable to expect the returns, variances, and Sharpe ratios for these three types of investors to have the following ex-ante relationships between the different types of investors (see Table~\ref{tab231}). In Table \ref{tab231}, the phrases `Around Large' and 'Around Small' are used with a slight abuse of notation to describe the nature of indeterminate relationships in expected returns between different investor types (Type-I and Type-U). For example, `Around Large' refers to the situation where, due to ESG uncertainty, it is unclear whether Type-U investors will receive higher returns than Type-I investors. This indeterminacy arises from the earlier proof, which showed that the relationship between Type-I and Type-U investors depends on the compensation for uncertainty relative to the potential loss from ESG integration. The results in Table \ref{tab231} reflect the trade-offs and uncertainties introduced by ESG considerations in portfolio decisions. For ESG-indifferent investors, the lack of constraints allows them to achieve high returns, low variance, and strong risk-adjusted performance (high Sharpe Ratios). In contrast, investors integrating ESG factors without uncertainty face reduced returns and diversification limitations, leading to higher variance and lower Sharpe Ratios. When ESG uncertainty is introduced, performance metrics become indeterminate: `Around Large' and `Around Small' highlight the fuzziness of the relationships among the return, variance, and Sharpe Ratio due to the ongoing trade-offs between potential compensation for managing uncertainty and the costs of ESG constraints. This also underscores the importance of addressing ESG rating inconsistencies in ensuring investors with better predictable outcomes.

\subsection{Calibration Exercise}


To assess the alignment between theoretical models and real-world portfolio selection outcomes, we conduct the following calibration exercises using the RL framework. In our calibration exercises, we treat ESG scores as representative proxies for firm-level sustainability, a practice widely adopted in both academic and institutional settings. ESG ratings are designed to reflect a firm’s exposure to and management of environmental, social, and governance risks. While they offer a structured lens into corporate responsibility, their validity as exact measures of “true” sustainability is inherently limited.

ESG scores differ across providers due to variations in methodologies, weighting schemes, and indicator selection. Additionally, many ESG metrics rely heavily on company self-disclosures, making them susceptible to selective reporting and data omissions—commonly referred to as disclosure bias (\cite{berg2022aggregate}). These limitations are further compounded by inconsistencies in coverage, especially across small-cap, private, or emerging market firms. Hence, ESG scores may reflect perceived sustainability more than objectively verifiable environmental or social impact.

Nevertheless, ESG ratings remain among the most accessible and standardized tools currently available for sustainability-oriented investment strategies. Their widespread use across asset managers, index providers, and regulatory bodies (e.g., the SFDR in the EU) supports their role as practical proxies for aligning capital allocation with sustainability goals. In this paper, we acknowledge the limitations of ESG ratings while recognizing their operational utility in empirical portfolio modeling. Accordingly, interpretations of the results should be framed within this context of imperfect but industry-relevant measurement.

\textbf{Data Source:} The study utilizes a dataset covering the period from 06/30/2018 to 06/30/2021 for the training set and 07/01/2021 to 06/30/2022 for the testing set. In each calibration, we choose to use ESG data from the four major ESG raters separately. In addition, the four kinds of proposed ensemble ESG scores in Section \ref{2.3} present an alternative ESG feature option for investors, and consequently, we conduct calibration exercises by incorporating the four suggested ensemble scores individually. This exploration allows us to comprehensively assess the viability and potential impact of these ensemble scores within our analytical framework. By considering these ensemble scores alongside individual ESG metrics, we endeavor to provide investors with a more nuanced understanding of their investment choices and their alignment with sustainable practices.

\textbf{RL Framework:} We maintain the configuration of the RL framework in accordance with the FinRL framework, as elaborated in Section \ref{2.2}. Our framework continues to comprise three distinct layers. Within the Agent layer, we still utilize the DDPG algorithm as the central agent for our FinRL-based investigations. In the bottom layer, the environment layer, we persist in simulating authentic financial markets, relying on historical data that encompasses various metrics, including closing prices, stock quantities, trading volume, and technical indicators.

Within the application layer, we integrate an algorithmic trading strategy into the RL framework by defining the state space, action space, and reward function. The state space remains inclusive of information such as account balance, current asset holdings, OHLC prices, trading volume, and technical indicators. The action space outlines the allowable actions through which the agent engages with the environment. Typically, an action is composed of three components: {-1, 0, 1}, where -1, 0, and 1 correspond to selling, holding, and purchasing one share, respectively. In terms of the reward function modification, the study explicitly incorporates the MV preference as a distinct reward function in the RL framework. This modified reward function is subsequently trained separately for the three types of investors we have specified earlier. To be specific, the reward function for each type of investor is set as:
\begin{itemize}
    \item Type-I Investors:
    \begin{equation}
        w \cdot \mu_M+(1-w) \mu_f-\frac{\gamma}{2} w^2 \sigma_M^2
    \end{equation}
    \item Type-N Investors: 
    \begin{equation}
        w \cdot \mu_M+(1-w) \mu_f-\frac{\gamma}{2} w^2 \sigma_M^2 + bw \cdot \mu_g
    \end{equation}
    \item Type-U Investors: 
    \begin{equation}
        w \cdot \mu_M+(1-w) \mu_f-\frac{\gamma}{2} w^2 \sigma_M^2+b\left(w \cdot \mu_g-\frac{\theta}{2} w^2 \sigma_g^2\right)
    \end{equation}   
\end{itemize}

After constructing the layers, we proceed to train the process by using FinRL for each calibration iteration. This training involves monitoring price fluctuations, making decisions (such as selling, holding, or buying a specific quantity of stock), and determining rewards based on various investor profiles. The outcomes generated by the models, for the three types of investors, utilizing ESG scores sourced from four individual ESG raters and four ensemble methods, are presented in Table \ref{tab232} and Table \ref{tab2322} below. The RL framework's configuration remains consistent across the creation of these two tables, with the only variation being the source of ESG scores or the ensemble methods applied. Both tables display the calibration results for three distinct investor types.

\begin{table}
\small 
\centering
\begin{tabular}{|p{3.0cm}|p{1.2cm}|p{0.8cm}|p{1.2cm}|p{0.8cm}|p{1.2cm}|p{0.8cm}|p{1.2cm}|p{0.8cm}|p{1.2cm}|p{0.8cm}|}
\hline
& \multicolumn{2}{|c|}{\textbf{$b<1(b=0.2)$}} & \multicolumn{2}{c|}{ \textbf{$b<1(b=0.6)$}} & \multicolumn{2}{c|}{ \textbf{$b=1$} } & \multicolumn{2}{c|}{ $b>1(b=1.4)$ } & \multicolumn{2}{c|}{ $b>1(b=1.8)$ } \\
\cline { 2 - 11 }
& \textbf{Return} & \textbf{SR} & \textbf{Return} & \textbf{SR} & \textbf{Return} & \textbf{SR} & \textbf{Return} & \textbf{SR} & \textbf{Return} & \textbf{SR} \\
\hline
\hline
ESG Indifference & $8.63\%$ & 0.77 &$8.63\%$ & 0.77 & $8.63\%$ & 0.77 & $8.63\%$ & 0.77 & $8.63\%$ & 0.77 \\
\hline
RebecoSAM (N) & $9.52\%$ & 0.78 & $8.01\%$ & 0.70 & $7.99\%$ & 0.66 & $7.93\%$ & 0.65 & $7.25\%$ & 0.53 \\
\hline
Sustain (N) & $7.66\%$ & 0.63 & $7.63\%$ & 0.66 & $7.68\%$ & 0.60 & $7.63\%$ & 0.58 & $7.45\%$ & 0.54 \\
\hline
MSCI (N) & $8.04\%$ & 0.72 & $7.89\%$ & 0.69 & $7.21\%$ & 0.58 & $6.26\%$ & 0.53 & $7.00\%$ & 0.52 \\
\hline
Asset4 (N) & $9.64\%$ & 0.81 & $8.21\%$ & 0.72 & $8.23\%$ & 0.70 & $8.03\%$ & 0.67 & $7.45\%$ & 0.55 \\
\hline
RebecoSAM (U) & $9.56\%$ & 0.77 & $8.45\%$ & 0.74 & $7.61\%$ & 0.68 & $8.31\%$ & 0.72 & $7.38\%$ & 0.47 \\
\hline
Sustain (U) & $9.61\%$ & 0.80 & $8.78\%$ & 0.81 & $8.54\%$ & 0.77 & $7.73\%$ & 0.62 & $7.51\%$ & 0.53 \\
\hline
MSCI (U) & $9.99\%$ & 0.88 & $8.42\%$ & 0.74 & $7.87\%$ & 0.69 & $7.62\%$ & 0.62 & $7.40\%$ & 0.58 \\
\hline
Asset4 (U) & $9.23\%$ & 0.84 & $8.63\%$ & 0.75 & $8.33\%$ & 0.72 & $8.23\%$ & 0.64 & $7.33\%$ & 0.48 \\
\hline
\end{tabular}
\centering
\caption{Calibration Results for Four Individual ESG Raters}
\label{tab232}
\end{table}

\begin{table}
\small 
\centering
\begin{tabular}{|p{3.0cm}|p{1.2cm}|p{0.8cm}|p{1.2cm}|p{0.8cm}|p{1.2cm}|p{0.8cm}|p{1.2cm}|p{0.8cm}|p{1.2cm}|p{0.8cm}|}
\hline
& \multicolumn{2}{|c|}{\textbf{$b<1(b=0.2)$}} & \multicolumn{2}{c|}{ \textbf{$b<1(b=0.6)$}} & \multicolumn{2}{c|}{ \textbf{$b=1$} } & \multicolumn{2}{c|}{ $b>1(b=1.4)$ } & \multicolumn{2}{c|}{ $b>1(b=1.8)$ } \\
\cline { 2 - 11 }
& \textbf{Return} & \textbf{SR} & \textbf{Return} & \textbf{SR} & \textbf{Return} & \textbf{SR} & \textbf{Return} & \textbf{SR} & \textbf{Return} & \textbf{SR} \\
\hline
\hline
ESG Indifference & $8.63\%$ & 0.77 &$8.63\%$ & 0.77 & $8.63\%$ & 0.77 & $8.63\%$ & 0.77 & $8.63\%$ & 0.77 \\
\hline
Centroid (N) & $8.45\%$ & 0.68 & $8.12\%$ & 0.65 & $7.38\%$ & 0.64 & $7.05\%$ & 0.64 & $6.68\%$ & 0.60 \\
\hline
Median (N) & $7.66\%$ & 0.63 & $7.63\%$ & 0.66 & $7.68\%$ & 0.62 & $7.63\%$ & 0.64 & $7.45\%$ & 0.63 \\
\hline
PCA (N) & $8.04\%$ & 0.72 & $7.89\%$ & 0.69 & $7.21\%$ & 0.66 & $6.26\%$ & 0.63 & $7.00\%$ & 0.62 \\
\hline
Alpha-Maxmin (N) & $9.03\%$ & 0.74 & $8.34\%$ & 0.70 & $8.12\%$ & 0.68 & $7.88\%$ & 0.65 & $7.43\%$ & 0.63 \\
\hline
Centroid (U) & $9.29\%$ & 0.73 & $8.62\%$ & 0.69 & $8.03\%$ & 0.64 & $7.68\%$ & 0.59 & $7.12\%$ & 0.53 \\
\hline
Median (U) & $9.61\%$ & 0.80 & $8.78\%$ & 0.81 & $8.54\%$ & 0.77 & $7.73\%$ & 0.62 & $7.51\%$ & 0.53 \\
\hline
PCA (U) & $9.23\%$ & 0.84 & $8.63\%$ & 0.75 & $8.33\%$ & 0.72 & $8.23\%$ & 0.64 & $7.33\%$ & 0.48 \\
\hline
Alpha-Maxmin (U) & $9.99\%$ & 0.88 & $8.72\%$ & 0.78 & $8.63\%$ & 0.72 & 7.62\% & 0.62 & $7.40 \%$ & 0.58 \\

\hline
\end{tabular}
\centering
\caption{Calibration Results for Four Ensemble ESG Scores}
\label{tab2322}
\end{table}

The results presented in Table \ref{tab232} and Table \ref{tab2322} can be summarily stated as follows. In scenarios without uncertainty, investors exhibiting indifference towards ESG factors typically achieve higher expected returns compared to those who prioritize ESG considerations. Conversely, investors who account for ESG uncertainty tend to realize higher expected returns than their counterparts who do not factor in such uncertainties. It is noteworthy to point out, however, that the expected returns and corresponding Sharpe ratios of investors with ESG uncertainty do not consistently surpass those of ESG-indifferent investors. Moreover, an increase in the parameter $b$ corresponds to a heightened emphasis on non-pecuniary returns, ultimately resulting in lower overall expected returns. Interestingly, the highest expected returns and Sharpe ratios are often attained by investors displaying indifference towards ESG factors or those considering ESG uncertainty. Notably, among the ensemble ESG score methodologies, Alpha-Maxmin consistently demonstrates superior performance in terms of expected portfolio return and Sharpe ratio when compared to other ensemble methods. While our empirical results demonstrate this consistency, we offer the following as a conjectured explanation for the performance advantage observed.

Unlike the other methods-such as PCA, which relies on statistical variance structure; the centroid (mean), which assumes equal informativeness across sources; or the median, which is robust to outliers but discards dependency information-the Alpha-Maxmin strategy directly incorporates investor preferences under ESG ambiguity. Specifically, it combines the best-case and worst-case ESG views using a tunable parameter $\alpha$. In our experiments, we set $\alpha=0.5$, representing an ambiguity-neutral investor who treats upside and downside ESG signals symmetrically. This structure may offer performance benefits in contexts of high ESG rating dispersion. The Alpha-Maxmin formulation effectively hedges against rating noise by preserving exposure to both favorable and unfavorable scenarios, reducing over-reliance on any single provider. This is especially useful when different raters embed different measurement scopes or methodologies, as documented in Berg et al. (2022). In contrast, purely data-driven methods such as PCA may be sensitive to correlation noise or data overfitting, while centroid and median lack a mechanism to express preference uncertainty. The relatively strong Sharpe ratios observed for Alpha-Maxmin under both Type-N and Type-U investor specifications may thus reflect its flexibility in adapting to both ESG return relevance and ESG score volatility. In particular, under Type-U preferences where ESG uncertainty is penalized directly, the Alpha-Maxmin strategy—already structured to internalize ambiguity—may align more naturally with the investor's objective function. While this reasoning remains a hypothesis grounded in observed patterns rather than formal theoretical derivation, it highlights a potentially meaningful structural advantage of Alpha-Maxmin that may warrant further theoretical and empirical investigation. Due to time and space constraints, we leave a full sensitivity analysis of the Alpha-Maxmin strategy to future work. In particular, it would be valuable to investigate how portfolio outcomes respond to variations in the ambiguity parameter $\alpha$, which governs the trade-off between worst-case and best-case ESG assessments. While this paper adopts the neutral case of $\alpha=0.5$ in the main calibration, future research could explore the full range from $\alpha=0$ (pure optimism) to $\alpha=1$ (pure pessimism) to assess the robustness of performance under different ambiguity attitudes and to uncover any asymmetries in return-risk tradeoffs driven by investor preferences.

The results presented above are broadly in agreement with the theoretical expectations of the model. However, there are a few isolated instances, in which the empirical results deviate from the theoretical predictions. Such deviations may arise due to a number of reasons, such as overfitting the model, a limited amount of the data used for training the model, etc.

The calibration study reveals that incorporating ESG considerations into investment decisions generally leads to a lower short-term financial performance, as evidenced by reduced Sharpe Ratios in the one-year test set. This finding suggests that in the short term, the focus on ESG may come at the expense of financial returns. However, in the long term, companies with a strong ESG performance are better positioned to manage policy or sustainability risks, potentially enhancing their financial resilience and performance over time. Additionally, we observe that the performance difference between Type-I and Type-U investors remains ambiguous. This reflects the trade-off between the loss of returns due to ESG considerations and the compensation for uncertainty through different ESG ratings.

The results of our calibration study, combined with the findings reported in the existing literature, underscore the complex and context-dependent relationship between ESG performance and financial returns. While some studies suggest that higher ESG performance can enhance financial outcomes, others argue that it may lead to reduced returns under specific conditions. For instance, \cite{does_pay_ESG2022} demonstrate that strong ESG practices are associated with better financial performance and higher market valuations in the U.S. market, where regulatory frameworks and consumer preferences increasingly favor sustainability. Similarly, \cite{MSCI_ESG2023} highlight the positive impact of ESG ratings on stock price performance, attributing this to reduced risk exposure and improved corporate resilience. Conversely, other research points to the variability in ESG's impact across different markets and contexts. For instance, \cite{esg_features2022} show that ESG scores do not uniformly translate to higher financial returns across sectors or company sizes in the European equity market, suggesting that sector-specific dynamics and investor priorities influence outcomes. Additionally, ESG considerations may introduce reputational risks, as evidenced by \cite{nicolas2024esg}, where spikes in ESG-related social media discussions led to significant reductions in abnormal returns, particularly in the Social and Governance pillars of ESG.

The lack of consensus in reaching clear conclusions in the literature can be attributed to several factors in their findings. First, the time horizon of analysis plays a crucial role. As our calibration study shows, in the short term, integrating ESG factors can reduce financial returns, particularly when ESG investments are priced at a premium due to a high demand. However, over the long term, companies with a strong ESG performance are better positioned to manage sustainability risks and adapt to regulatory changes, potentially enhancing resilience and generating superior returns. Second, regional differences in ESG awareness and policy adoption are significant contributors to these discrepancies. For instance, countries with well-established regulatory frameworks and strong investor interest in ESG may exhibit stronger positive correlations between ESG and financial performance compared to regions where ESG is less prominently emphasized. These regional differences highlight how government policies and cultural attitudes towards sustainability can shape market outcomes. Third, variations in the time periods of the datasets used in these studies further explain the discrepancies. Research conducted in the early stages of ESG adoption, when awareness and integration were limited, may show weaker or even negative correlations between ESG and financial performance. In contrast, more recent studies, conducted at a time when ESG factors are more widely acknowledged and integrated into the investment decision-making process, tend to demonstrate stronger positive impacts. This shift reflects the growing market acceptance of ESG as a critical component of corporate performance. Finally, differences in the scope and weighting of ESG components across studies also contribute to varying results. For example, studies emphasizing environmental factors may show stronger correlations with financial performance in industries such as renewable energy, while studies focusing on social or governance factors might yield mixed results.

These findings reinforce the need for investors to align their ESG preferences with their investment goals and risk tolerance. They also highlight the importance of addressing ESG rating heterogeneity and ensuring that long-term sustainability objectives are routinely factored into portfolio management strategies.

\section{ESG-modified Capital Asset Pricing Model}
\label{2.5}

For completeness of our analysis, we extend the Capital Asset Pricing Model (CAPM) introduced by 
\cite{sharpe1964capital} to incorporate ESG scores and account for uncertainty, leveraging the results previously obtained from the MV model. The CAPM has long been a cornerstone of modern finance, providing a theoretical framework for understanding the trade-off between risk and return. Traditionally, the CAPM assumes that all investors consider market risk as the sole factor affecting expected returns. However, with the growing emphasis on ESG considerations in investment decision-making, researchers have sought to expand the CAPM framework to account for these factors. Recent literature explores various methods of incorporating ESG constraints into the traditional model. For instance, some studies focus on adjusting the utility function to include ESG preferences, thereby redefining the efficient frontier to reflect sustainable investment priorities \cite{pedersen2021responsible}. Other works examine the relationship between corporate social responsibility and the cost of equity, highlighting how strong ESG performance reduces risk premiums for firms \cite{albuquerque2019corporate}. Additionally, meta-analyses and empirical research reveal that ESG factors are increasingly regarded as non-diversifiable risks that should be incorporated into asset pricing models \cite{friede2015meta}.

Despite these advancements, the current literature often assumes that ESG factors are either deterministic or can be linearly incorporated into the model. These assumptions overlook the inherent uncertainty in the ESG ratings, which often vary significantly across the different rating agencies. Moreover, many existing models treat the ESG factors as static inputs, failing to address the dynamic nature of the ESG considerations in response to market conditions or investor preferences. The CAPM model proposed in this section seeks to address these limitations by introducing a novel Double Mean-Variance framework that incorporates ESG constraints while explicitly accounting for uncertainty. A key innovation of this framework is its ability to decouple the components of return, risk, and uncertainty aversion. This decomposition provides a more general formulation of the CAPM, allowing for a greater degree of flexibility in modeling investor behavior. For example, the framework separates the effects of the market return and the ESG return, enabling independent adjustments of risk and ESG uncertainty. This separation is particularly advantageous for investors with diverse preferences or regulatory constraints, as it allows for tailored adjustments to portfolio weights based on specific aversion levels.

To build up an ESG-CAPM, firstly, we provide a list of variable declarations in Table \ref{tab3}.

\begin{table}[H]
\begin{tabular}{|l|l|l|}
\hline
\textbf{Variable} & \textbf{Description} & \textbf{Dimension} \\
\hline
$\boldsymbol{X}_i$ & Agent $i$'s portfolio weights for $n$ stocks & $n \times 1$ vector\\
$\boldsymbol{X}_M$ & Market portfolio vector & $n \times 1$ vector\\
$\boldsymbol{\mu}_r$ & Market excess return vector & $n \times 1$ vector\\
${\mu}_M=\boldsymbol{X}_M^T \boldsymbol{\mu}_r$ & Market expected return & A number\\
$\boldsymbol{\mu}_{g,M}$ & Expected ESG score vector & $n \times 1$ vector\\
${\mu}_g = \boldsymbol{X}_M^T\boldsymbol{\mu}_{g,M}$ & Market ESG score & A number\\
$\boldsymbol{\Sigma}_{M}$ & Covariance matrix of returns & $n \times n$ matrix\\
$\boldsymbol{\Sigma}_{g,M}$ & Covariance matrix of ESG ratings & $n \times n$ matrix\\
$w_i$ & Weight of Agent $i$ & A number\\
$\sigma_M^2=\boldsymbol{X}_M^{T} \boldsymbol{\Sigma}_M \boldsymbol{X}_M$ & Market return variance & A number\\
$\sigma_{g}^2=\boldsymbol{X}_M^{T} \boldsymbol{\Sigma}_{g,M} \boldsymbol{X}_M$ & Market ESG score variance & A number\\
\hline
\end{tabular}
\centering
\caption{Description of Variables}
\label{tab3}
\end{table}


The direct utility function for investor $i$ is formulated as:
\begin{equation}
\boldsymbol{X}_i^{T} \boldsymbol{\mu}_r-\frac{\gamma_i}{2} \boldsymbol{X}_i^{T} \boldsymbol{\Sigma}_{M} \boldsymbol{X}_i+b_i\left(\boldsymbol{X}_i^{T} \boldsymbol{\mu}_{g, M}-\frac{\theta_i}{2} \boldsymbol{X}_i^{T} \boldsymbol{\Sigma}_{g,M} \boldsymbol{X}_i\right)
\end{equation}

After taking the first-order condition, the agent $i$'s portfolio weights for $n$ stocks can be expressed as:
\begin{align}
\boldsymbol{X}_i &=\left(
\gamma_i \boldsymbol{\Sigma}_{M}+b_i \theta_i \boldsymbol{\Sigma}_{g,M}\right)^{-1}\left(\boldsymbol{\mu}_r+b_i \boldsymbol{\mu}_{g, M}\right) \nonumber \\
&=\left(\gamma_i \boldsymbol{\Sigma}_{M}+b_i \theta_i \boldsymbol{\Sigma}_{g,M}\right)^{-1} \cdot \gamma_i \boldsymbol{\Sigma}_{M}\left(\frac{1}{\gamma_i} \boldsymbol{\Sigma}_{M}{ }^{-1}\right) \cdot\left(\boldsymbol{\mu}_r+b_i \boldsymbol{\mu}_{g, M}\right) \nonumber \\
&=\left(\gamma_i \boldsymbol{\Sigma}_{M}+b_i \theta_i \boldsymbol{\Sigma}_{g,M}\right)^{-1} \cdot \gamma_i \boldsymbol{\Sigma}_M \cdot \frac{1}{\gamma_i}\left[\boldsymbol{\Sigma}_{M}^{-1}\left(\boldsymbol{\mu}_r\right)+\boldsymbol{\Sigma}_{M}{ }^{-1} b_i \boldsymbol{\mu}_{g, M}\right] \nonumber \\
&={\left[\frac{1}{\gamma_i} \boldsymbol{\Sigma}_{M}^{-1}\left(\boldsymbol{\mu}_r\right)+\frac{1}{\gamma_i} \boldsymbol{\Sigma}_{M}{ }^{-1} b_i \boldsymbol{\mu}_{g, M}\right]} \nonumber \\
&-{\left[\left(\gamma_i \boldsymbol{\Sigma}_{M}+b_i \theta_i \boldsymbol{\Sigma}_{g,M}\right)^{-1} \cdot b_i \theta_i \boldsymbol{\Sigma}_{g,M}\right]{\left[\frac{1}{\gamma_i} \boldsymbol{\Sigma}_{M}^{-1}\left(\boldsymbol{\mu}_r\right)+\frac{1}{\gamma_i} \boldsymbol{\Sigma}_{M}{ }^{-1} b_i \boldsymbol{\mu}_{g, M}\right]}}
\label{equ1}
\end{align}
\subsection{ESG-modified CAPM without uncertainty} 

In the absence of ESG uncertainty, we only take the first two terms of Equation (\ref{equ1}) into consideration, and as a result, the market portfolio is given by:
\begin{equation}
\boldsymbol{X}_M=\sum_{i=1}^I w_i\left[\frac{1}{\gamma_i} \boldsymbol{\Sigma}_{M}^{-1}\left(\boldsymbol{\mu}_r\right)+\frac{1}{\gamma_i} b_i \cdot \boldsymbol{\Sigma}_{M}{ }^{-1} \cdot \boldsymbol{\mu}_{g, M}\right]
\end{equation}
Then, 
\begin{equation}
\boldsymbol{\Sigma}_{M} \boldsymbol{X}_M=\sum_{i=1}^I \frac{w_i}{\gamma_i} \cdot \boldsymbol{\mu}_r+\sum_{i=1}^I \frac{w_i}{\gamma_i} \cdot b_i \cdot \boldsymbol\mu_{g, M}
\end{equation}
Let $\gamma_M=\frac{1}{\sum_{i=1}^I w_i \cdot \frac{1}{\gamma_i}}$ and $b_M=\frac{\sum_{i=1}^I w_i \cdot \frac{b_i}{\gamma_i}}{\sum_{i=1}^I w_i \cdot \frac{1}{\gamma_i}}=\left(\sum_{i=1}^I w_i \cdot \frac{b_i}{\gamma_i}\right) \cdot \gamma_M$, then 
\begin{equation}
\boldsymbol{\Sigma}_{M} \boldsymbol{X}_M=\frac{1}{\gamma_M} \cdot \boldsymbol\mu_r+\frac{b_M}{\gamma_M} \cdot \boldsymbol\mu_{g, M},
\label{equ7}
\end{equation}

\begin{equation}
\begin{aligned}
\boldsymbol\mu_r & =\gamma_m \boldsymbol{\Sigma}_M \boldsymbol{X}_M-b_M \cdot \boldsymbol\mu_{g, M} \\
& =\frac{\boldsymbol{\Sigma}_{M} \boldsymbol{X}_M}{\sigma_M^2} \cdot \gamma_M \sigma_M^2-b_M \cdot \boldsymbol\mu_{g, M}
\label{equ2}
\end{aligned}
\end{equation}
Next, we derive the market expected return ${\mu}_M$ by multiplying $\boldsymbol{X}_M^T$ on both sides of Equation (\ref{equ2}).
\begin{equation}
\mu_M=\boldsymbol{X}_M^{T} \boldsymbol{\mu}_r=\gamma_M \boldsymbol{X}_M^{T} \boldsymbol{\Sigma}_{M} \boldsymbol{X}_M-\boldsymbol{X}_M^{T}\left(b_M \cdot \boldsymbol\mu_{g, M}\right) = \gamma_M\sigma_M^2-\boldsymbol{X}_M^{T}\left(b_M \cdot \boldsymbol\mu_{g, M}\right)
\label{equ3}
\end{equation}
From Equation (\ref{equ3}), we know $\gamma_M\sigma_M^2 = \mu_M + \boldsymbol{X}_M^{T}\left(b_M \cdot \boldsymbol\mu_{g, M}\right)$. Then, we replace $\gamma_M\sigma_M^2$ by $\mu_M + \boldsymbol{X}_M^{T}\left(b_M \cdot \boldsymbol\mu_{g, M}\right)$ in Equation (\ref{equ2}). 

After re-arranging terms, an ESG-modified CAPM without uncertainty is shown in Equation (\ref{equ4}), where $\boldsymbol{\beta} = \frac{\boldsymbol{\Sigma}_{M} \boldsymbol{X}_M}{\sigma_M^2}$. Here, Alpha ($\boldsymbol{\alpha}$) can be expressed as $b_M\left(\boldsymbol\beta \mu_g-\boldsymbol\mu_{g, M}\right)$.
\begin{equation}
\begin{aligned}
\boldsymbol\mu_r &=\frac{\boldsymbol{\Sigma}_{M} \boldsymbol{X}_M}{\sigma_M^2} \cdot \mu_M+\frac{\boldsymbol{\Sigma}_{M} \boldsymbol{X}_M}{\sigma_M^2} \boldsymbol{X}_M{ }^{T} \cdot b_M \cdot \boldsymbol\mu_{g, M}-b_M \cdot \boldsymbol\mu_{g, M} \\
& = \boldsymbol\beta \mu_M+b_M\left[\boldsymbol\beta \mu_g-\boldsymbol\mu_{g, M}\right]
\label{equ4}
\end{aligned}
\end{equation}
The expected excess return of investors is affected by two factors, which are a modified market premium (purely market features $\boldsymbol\beta \mu_M$) as well as an Alpha component (features related to ESG scores $b_M\left[\boldsymbol\beta \mu_g-\boldsymbol\mu_{g, M}\right]$). The alpha value, which represents excess returns not explained by the stock's beta and the market return, is influenced by the difference between a company's own ESG score and the market ESG score, multiplied by the stock's beta. 

To briefly illustrate this concept by means of a numerical example, consider a stock with a beta of 1.5 and a market ESG score of 2. If the stock's own ESG score is 2.5, it will have a positive alpha ($+0.5$). On the other hand, if the ESG score is greater than 3.0, the alpha will be negative. This also is in line with the previous conclusion, i.e., after considering the preference for the ESG factor, the expected return will decrease. A higher ESG requirement will lead to a greater loss of expected return.

\subsection{ESG-modified CAPM with uncertainty} 

For the case with ESG uncertainty, we should consider the full form of Equation (\ref{equ1}) as the agent $i$’s portfolio weights for $n$ stocks. Thus, the market portfolio is given by:

\begin{equation}
\begin{aligned}
\boldsymbol{X}_M & =\sum_{i=1}^I w_i \cdot \boldsymbol{X}_i \\
& =\sum_{i=1}^I w_i \cdot\left(\gamma_i \boldsymbol{\Sigma}_{M}+b_i \theta_i \boldsymbol{\Sigma}_{g,M}\right)^{-1} \cdot \boldsymbol{\Sigma}_{M}^{-1}\left(\boldsymbol\mu_r+b_i \boldsymbol\mu_{g, M}\right)
\end{aligned}
\end{equation}
\begin{equation}
\begin{aligned}
\boldsymbol{\Sigma}_{M} \boldsymbol{X}_M&=\sum_{i=1}^I w_i\left(\gamma_i \boldsymbol{\Sigma}_{M}+b_i \theta_i \boldsymbol{\Sigma}_{g,M}\right)^{-1}\left(\boldsymbol\mu_r+b_i \boldsymbol\mu_{g, M}\right) \\
&=\sum_{i=1}^I w_i\left(\gamma_i \boldsymbol{\Sigma}_{M}+b_i \theta_i \boldsymbol{\Sigma}_{g,M}\right)^{-1} \boldsymbol\mu_r+\sum_{i=1}^I w_i\left(\gamma_i \boldsymbol{\Sigma}_{M}+b_i \theta_i \boldsymbol{\Sigma}_{g,M}\right)^{-1}  b_i \cdot\boldsymbol\mu_{g, M} \\
&={\boldsymbol\Gamma_{M,U}}^{-1} \cdot \boldsymbol\mu_r+{\boldsymbol{B}_{M,U}}{\boldsymbol{\Gamma}_{M,U}}^{-1} \cdot \boldsymbol\mu_{g, M}
\label{equ6}
\end{aligned}
\end{equation}
where $\boldsymbol\Gamma_{M,U} = {\left(\sum_{i=1}^I w_i\left(\gamma_i \boldsymbol{\Sigma}_{M}+b_i \theta_i \boldsymbol{\Sigma}_{g,M}\right)^{-1}\right)}^{-1}$,
and 
\begin{equation}
\begin{aligned}
\boldsymbol{B}_{M,U} & = {\left(\sum_{i=1}^I w_i\left(\gamma_i \boldsymbol{\Sigma}_{M}+b_i \theta_i \boldsymbol{\Sigma}_{g,M}\right)^{-1}  b_i\right)} \cdot {\left(\sum_{i=1}^I w_i\left(\gamma_i \boldsymbol{\Sigma}_{M}+b_i \theta_i \boldsymbol{\Sigma}_{g,M}\right)^{-1}\right)}^{-1} \\
& =\boldsymbol{\Gamma}_{M,U} \left(\sum_{i=1}^I w_i\left(\gamma_i \boldsymbol{\Sigma}_{M}+b_i \theta_i \boldsymbol{\Sigma}_{g,M}\right)^{-1}  b_i\right).
\end{aligned}
\end{equation}
Equation (\ref{equ6}) has the same form as Equation (\ref{equ7}). The following derivation process should be the same as the case without ESG uncertainty. Thus, the CAPM for Type-U investors is given by:
\begin{equation}
\begin{aligned}
\boldsymbol\mu_r &=\frac{\boldsymbol{\Sigma}_{M} \boldsymbol{X}_M}{\sigma_M^2} \cdot \mu_M+\frac{\boldsymbol{\Sigma}_{M} \boldsymbol{X}_M}{\sigma_M^2} \boldsymbol{X}_M{ }^{T} \cdot \boldsymbol{B}_{M,U} \cdot \boldsymbol\mu_{g, M}-\boldsymbol{B}_{M,U} \cdot \boldsymbol\mu_{g, M} \\
& = \boldsymbol\beta \mu_M+\boldsymbol{B}_{M,U}\left(\boldsymbol\beta \mu_g-\boldsymbol\mu_{g, M}\right),
\label{equ42}
\end{aligned}
\end{equation}
where $\boldsymbol{\beta} = \frac{\boldsymbol{\Sigma}_{M} \boldsymbol{X}_M}{\sigma_M^2}$ and the Alpha ($\boldsymbol{\alpha}$) can be expressed as $\boldsymbol{B}_{M,U}\left(\boldsymbol\beta \mu_g-\boldsymbol\mu_{g, M}\right)$.

The final form for Type-U investors is the same as the case for the type-N investors, except that the $\boldsymbol{B}_{M,U}$ is defined with the incorporation of $\boldsymbol{\Sigma}_{g,M}$ to express the features of ESG Uncertainty. Similarly, when dealing with situations that encompass uncertainty, the Alpha-value retains its sensitivity to the disparity between a company's specific ESG score and the market's ESG score, multiplied by the stock's beta. Employing the same example as previously demonstrated, envision a stock with a beta of 1.5 and a market ESG score of 2. In this context, if the stock's individual ESG score remains at 2.5, a positive alpha ($+0.5$) will be generated. Conversely, should the ESG score exceed 3.0, the resultant alpha will shift to a negative value.

In principle, a full calibration exercise can be carried out to illustrate the efficacy of the ESG-modified CAPM with and without uncertainty and the results can then be compared with those obtained from the traditional CAPM. This is beyond the scope of the present study and left for a future study.

\section{Conclusions}
\label{2.6}


Modern investors prioritize ESG scores, aligning their investments with companies that champion environmental responsibility, social awareness, and robust governance. The landscape of ESG data providers is diversified and encompasses well-established players. The existing literature tends to focus on the beneficial role of ESG integration in generating expected excess returns or risk premia. 

With the abundance of variety comes the inevitability of heterogeneity. It is not uncommon for the rating agencies to disagree with their published ESG ratings for companies in their samples. This divergence in the ESG ratings affects both the investors, who wish to achieve the financial and ESG-factor return objectives, and the companies, who seek to improve their ESG performances. 

The heterogeneity of ESG ratings by major rating agencies is treated in this paper as a source of ambiguity. In response to this ambiguity, we propose four ESG ensemble strategies that cater to different risk and ambiguity preference profiles, rather than providing a single, optimal ESG scoring system. Our findings underscore the urgent need for standardizing ESG metrics to address the inconsistencies highlighted in this paper. By revealing the challenges posed by divergent ESG measures, this study contributes to the growing recognition that standardization and transparency are critical for the effective adoption of ESG investing. The proposed ensemble strategies can serve as a bridge until a generally acceptable framework is developed and embraced. Furthermore, the steps outlined earlier, including metric standardization, methodological transparency, and independent oversight, offer actionable pathways to achieve greater consistency in ESG measures. These efforts are not only feasible but also necessary to enhance the reliability of ESG metrics and their integration into financial decision-making processes.

Next, in our portfolio optimization strategy in Section \ref{2.2}, we explicitly incorporated ESG scores in the reward function of the RL model to analyze the effect of widespread confusion caused by the heterogeneity in ESG ratings on the coherence of investment strategies. 

To combine the goals of achieving both financial market return (pecuniary return) and ESG-factor return (non-pecuniary return) objectives, we presented a DMV model, a linear combination of the two objectives with a preference ratio parameter. In this model, we defined three types of investors: those who are indifferent to ESG, those who prefer ESG but are not affected by uncertainty with respect to the ESG ratings, and those who prefer ESG and are affected by uncertainty with respect to the ESG ratings. Our analysis of the optimal portfolio weights for these investors can be summarized as follows: (i) the no-uncertainty case is associated with a negative ESG incremental premium; (ii) uncertainty with respect to the ESG ratings increases the risk for the investors, and the required return reflects this added risk; and (iii) the sign of the incremental premium is indeterminate when comparing the ESG indifference case and the ESG preference with uncertainty case due to conflicting forces at work between the risk and the return of the ESG factor. 

Additionally, the relationships for the variances of the investors are reversed to those for the expected mean relationships. When incorporating ESG ratings with uncertainty in the model, its variance is larger than that without incorporating ESG ratings.  However, the difference between the ESG indifference case and the ESG preference with uncertainty case is indeterminate. To verify the practicality of the proposed theoretical framework, we use an RL model by setting the reward function to be the specified DMV model.

To complete our analysis, we leverage the results obtained for the DMV model to construct ESG-modified CAPMs for ESG preference investors with and without uncertainty with respect to ESG ratings, in order to evaluate the performance of the resulting optimized portfolio. 

Both models are shown to have a similar structure and divide the expected return into two components, which are the modified market premium (based on the purely market factors) and the Alpha component (based on the ESG factor). Our analysis showed that the Alpha value is affected by the difference between a company's individual ESG ratings and the market ESG ratings, multiplied by the stock's beta. This information can be used to determine a threshold for positive Alpha in a sustainable investment approach, aiming at shedding light on these intricacies, and guiding both investors and company managers towards better-informed sustainable choices. In principle, a full-fledged empirical study can be performed by using the analytical framework developed in this subsection based on real data. This is left for a future study.


The heterogeneity of ESG ratings, which arises from differences in scoring methodologies, indicator weights, data transparency, and reporting standards, remains a well-documented and persistent challenge in sustainable finance. While ESG ratings aim to reflect firm-level sustainability performance, they are far from standardized and often exhibit significant divergence. Given the current lack of a unified global ESG framework and the reliance on firm self-disclosure, this heterogeneity is unlikely to be resolved soon. Yet ESG ratings remain one of the most widely used tools by investors seeking to incorporate sustainability into their investment process.

In this context, the ensemble methods proposed in this paper serve as a practical and adaptive solution. Rather than forcing investors to choose among conflicting ESG providers, our framework enables them to integrate multiple signals in a structured, preference-sensitive manner. When an investor is unsure which provider to trust—or wishes to mitigate model-specific biases—ensemble strategies such as Alpha-Maxmin offer a principled fallback. The ambiguity parameter $\alpha$ allows users to tune the level of conservatism, making the method adaptable to diverse institutional mandates, regulatory regimes, and stakeholder expectations.

For institutional investors, the proposed ensemble approach can be implemented across various levels of the investment decision-making process. Quantitative portfolio managers may embed ensemble ESG scores directly into multi-factor models or use them as tilting variables in portfolio optimization to enhance robustness. In both long-only and long-short strategies, ensemble-based ESG signals can support ranking, screening, and exposure control, helping reduce over-reliance on any single rating provider. Risk and compliance teams can adopt ensemble scores as more stable inputs for stress testing, ESG policy overlays, or internal threshold setting—especially under evolving regulatory regimes such as the Sustainable Finance Disclosure Regulation (SFDR) and the Task Force on Climate-Related Financial Disclosures (TCFD). At the strategic level, investment committees and asset allocators may utilize ensemble ESG tools to evaluate relative performance across sectors, geographies, or asset classes, particularly in regions or industries where ESG ratings are known to be highly inconsistent. 

Beyond its theoretical contributions, this study also carries important practical implications. The ensemble methods proposed here—including both statistical and preference-weighted approaches—can be integrated into existing portfolio construction pipelines by asset managers and institutional investors. These methods can serve as preprocessing layers to aggregate ESG scores from multiple providers before feeding them into optimization or trading systems. In reinforcement learning-based settings, the integration is even more direct: our preference-adjusted ESG scores can be included in the agent’s state representation or reward design, depending on the investment objective and ESG mandate.

By allowing for flexible adaptation to ESG rating disagreement, these models enable investors to maintain exposure to sustainability-aligned assets while mitigating the noise and inconsistency inherent in current ESG data. This is especially valuable for long-term investors such as pension funds, sovereign wealth funds, or endowments that are increasingly subject to ESG accountability. Furthermore, widespread adoption of ensemble ESG integration may also influence market behavior by reducing overreliance on any single rating agency and improving the consistency of ESG-linked capital allocation across the financial system.

Overall, our ensemble approach provides a practical bridge between academic insights and real-world portfolio decisions in the face of persistent ESG uncertainty. It supports both robust financial performance and credible alignment with sustainability objectives.


In a nutshell, this paper provides a quantitative framework for handling ESG rating uncertainty in portfolio allocation with three major applications in finance, but several avenues for future research remain outstanding. First, our model classifies investors based on their risk preferences and tolerance toward ESG ambiguity. However, investor heterogeneity extends beyond these dimensions. Ethical priorities, regional regulatory requirements, and stakeholder mandates can vary significantly across jurisdictions and investor types. Future work could extend the DMV framework by allowing the ESG preference parameter to vary across agents or evolve dynamically to reflect these additional constraints.

Second, while we treat ESG scores as exogenous inputs, there may exist important feedback effects between investor behavior and ESG ratings. For example, sustained capital reallocation based on ESG scores could influence firms’ incentives to disclose information, improve governance, or engage in greenwashing. Modeling the interaction between ESG ratings and investor actions in a dynamic setting would provide a richer understanding of ESG-related market equilibrium.

Third, our empirical calibration focuses on large-cap U.S. equities. Applying the ensemble approach to emerging markets or to firms with less consistent ESG disclosure may reveal new challenges, particularly given their lower data quality and limited coverage. It would also be valuable to test the ensemble strategies under different macroeconomic regimes—such as during periods of inflation stress, monetary tightening, or geopolitical instability—to evaluate their robustness across diverse economic conditions.

Together, these suggested extensions may have the prospect of further enhancing the applicability of the framework proposed in this study in a variety of practical settings.

\newpage

\bibliographystyle{elsarticle-harv}
\bibliography{main}

\end{document}